\documentclass[prd,preprint,floatfix,nofootinbib,superscriptaddress,showpacs]{revtex4} 

\pdfoutput=1
\usepackage{epsfig}
\usepackage[caption=false]{subfig}
\usepackage{amsmath}
\usepackage{amsfonts}
\usepackage{amssymb}
\usepackage{float}
\usepackage{color}
\usepackage{graphicx}
\usepackage{graphics}
\usepackage{hyperref}
\hypersetup{}
\usepackage[utf8x]{inputenc} 
\usepackage{epstopdf}
\usepackage{multirow}
\usepackage{slashed}
\definecolor{rosso}{cmyk}{0,1,1,0.4}
\definecolor{rossos}{cmyk}{0,1,1,0.55}
\definecolor{rossoc}{cmyk}{0,1,1,0.2}
\definecolor{blu}{cmyk}{1,1,0,0.3}
\definecolor{blus}{cmyk}{1,1,0,0.6}
\definecolor{bluc}{cmyk}{1,1,0,0.1}
\definecolor{verde}{cmyk}{0.92,0,0.59,0.25}
\definecolor{verdec}{cmyk}{0.92,0,0.59,0.15}
\definecolor{verdes}{cmyk}{0.92,0,0.59,0.4}

\hypersetup{
 colorlinks=true,
 linkcolor=verdec,
 urlcolor=verdec,
 citecolor=verdec
}

\begin{document}

\title{\color{verdes} Searching for GeV-scale Majorana Dark Matter: inter spem et metum}

\author{Adil Jueid}
\email{adil.hep@gmail.com}
\address{Department of Physics, Konkuk University, Seoul 05029, Republic of Korea}

\author{Salah Nasri}
\email{salah.nasri@gmail.com}
\address{Department of physics, United Arab Emirates University, Al-Ain, UAE}
\address{The Abdus Salam International Centre for Theoretical Physics, Strada
Costiera 11, I-34014, Trieste, Italy}

\author{Rachik Soualah}
\email{rsoualah@sharjah.ac.ae}
\address{Department of Applied Physics and Astronomy,University of Sharjah, P. O. Box 27272 Sharjah, UAE}
\address{The Abdus Salam International Centre for Theoretical Physics, Strada
Costiera 11, I-34014, Trieste, Italy}

\begin{abstract}
We suggest a minimal model for GeV-scale Majorana Dark Matter (DM) coupled to the standard model lepton sector  via  a charged scalar singlet. We show that there is an anti-correlation between the spin-independent DM-Nucleus scattering cross section ($\sigma_{\mathrm{SI}}$) and the DM  relic  density for parameters values allowed by various theoretical and experimental constraints. Moreover, we find  that even when DM couplings  are  of order unity, $\sigma_{\mathrm{SI}}$ is below the current experimental bound  but above the neutrino floor. Furthermore, we show that the considered model can be probed at high energy lepton colliders using e.g. the mono-Higgs production and same-sign charged Higgs pair production.
\end{abstract}


\maketitle
\tableofcontents

\section{Introduction}
\label{sec:intro}

The existence of Dark Matter (DM) in the universe is an established fact  supported by various observations at the sub-galactic,  galactic, and  cosmological scales (for a review, see e.g. \cite{Bertone:2004pz}). The measurements of CMB anisotropies implied that DM constitutes about $80\%$ of the matter budget in the universe with density of $\Omega_\mathrm{DM} h^2 = 0.1198\pm0.0015$ \cite{Ade:2015xua}, and  the  standard theories of  structure formation requires it to be non-relativistic when gravitational clustering started at the matter-radiation equality.  This  type of DM is  known as  cold DM (CDM). One of  the simplest scenarios of CDM is the thermal-freeze out mechanism in which  the DM  can be  accommodated   by    Weakly Interacting Massive Particles (WIMPs)   produced in thermal bath and  as the universe expands and cools down their relic abundance freeze-out when  temperature drops below their mass. Possible evidence for  WIMP DM has driven numerous experimental efforts to search for it via  direct detection \cite{Aprile:2018dbl, Cui:2017nnn}, indirect detection \cite{Adriani:2010rc, Aguilar:2013qda, Ahnen:2016qkx, Abdallah:2016ygi}  and in  collider experiments \cite{Meirose:2016pxn, Ahuja:2018bbj}.  Unfortunately, no signal for CDM has been detected,  and consequently upper limit on the cross section of their scattering off nucleus in the mass range  GeV to TeV are obtained, with the stringent  bound  of $4.1 \times 10^{-47}~\text{cm}^2$ for a mass of $30~\text{GeV}$. It is expected that the upcoming XENONnT experiment will be able to probe cross sections smaller by more than an order of magnitude \cite{Aprile:2020vtw}. Moreover, a model-independent estimate of the limits on the  annihilation cross sections of WIMP DM  implies  strong constraints on models with $s$-wave annihilation channel \cite{Leane:2018kjk}. These strong exclusions combined with the null results from direct detection experiments put an end to the most minimal model for GeV-scale DM candidate, i.e. the SM with real scalar singlet \cite{Silveira:1985rk, Burgess:2000yq} which called for several extensions \cite{Arcadi:2016qoz, Casas:2017jjg}.

Models with a singlet Majorana fermion as a DM candidate can potentially avoid these constraints due to the fact that their annihilation is dominated by $p$-wave amplitudes (which are suppressed by the square of the DM velocity) in addition  of being both minimal and predictive. On the other hand, the scattering  of the Majorana DM off the nucleus is induced at the one-loop order due to the absence of tree level couplings of the Majorana DM to the $Z/H$ bosons. Consequently, models containing Majorana DM can evade easily direct detection constraints even for model parameters of order $\mathcal{O}(1)$ and  hence avoiding the over-abundance condition $\Omega_\mathrm{WIMP} h^2 \leqslant \Omega_{\mathrm{Planck}} h^2$. Models containing Majorana DM are phenomenologically very attractive as they can easily address the question of the smallness of neutrino mass through radiative mass generation mechanism, and the problem of baryon asymmetry in the universe (BAU) via electroweak baryogenesis \cite{Ma:1998dn, Krauss:2002px, Aoki:2008av, Gustafsson:2012vj, Boucenna:2014zba, Cai:2017jrq}. 
In this work, we  suggest a very simple model which extends the Standard Model (SM) particle content by two gauge singlets: a charged scalar and a Majorana fermion. Using simple correlations between the relic abundance and the predicted spin-independent cross section, we show that the model is neither excluded nor close to the neutrino-floor region. We study the potential discovery of the DM and the charged Higgs in both $e^-e^-$ and $e^-e^{+}$  processes. We find that it can be  exclusively probed at lepton colliders such as the international linear collider (ILC) while the current and future LHC constraints will play a complementary role.

The remainder of this paper is organized  as follows: In section \ref{sec:model}, we discuss briefly the model's new degrees of freedom and their possible interactions. In section \ref{sec:constraints}, we study the various theoretical and experimental constraints on  parameter space of the model; in particular, we study constraints from perturbative unitarity, boundness-from-below of the scalar potential, ${\rm BR}(H\to\gamma\gamma)$ measurements, and the impact from bounds from lepton-flavor violating decays and Higgs boson invisible decays. In section \ref{sec:LHC}, we discuss of the impact of searches of charginos/sleptons at the LHC on the parameter space of our model. In section \ref{sec:DM}, we discuss in detail the DM phenomenology of our model with emphasis on the relic abundance of DM candidate, and its correlation to the spin-independent DM-Nucleus cross section. The collider implications are discussed in section \ref{sec:ILC}. We summarize our results in section \ref{sec:conclusion}. Some technical details are reported in Appendices \ref{sec:appendix1} and \ref{sec:MA5}.

\section{The model}
\label{sec:model}

We consider a minimal extension of the SM 
which contains, in addition to the SM particle content, two gauge-singlet fields; 
a charged scalar $S$ and a right handed (RH) neutral fermion $N$. Under the SM gauge group  $SU(3) \otimes SU(2)_L \otimes U(1)_Y$, the two new degrees of freedom transform as 
\begin{eqnarray}
S : ({\bf{1}, 1, 2}), \qquad \textrm{ and } \qquad N : ({\bf{1}, 1, 0}).
\end{eqnarray}
Moreover, to ensure the stability of DM in the universe, we impose
an exact discrete symmetry $Z_2$ under which all the SM fields are even while the new states are odd, i.e. $\{V^\mu, \Phi, \ell, q\} \to \{V^\mu, \Phi, \ell, q \}$, and 
$\{S, N\} \to \{-S, -N \}$, with the condition that $N$ is always lighter than the charged scalar singlet. In this simple
model, the RH fermion plays the role of the DM candidate while the charged scalar 
is a mediator of the DM-visible interaction.  Under these symmetry requirements, the most general Yukawa Lagrangian can be written as\footnote{$h.c = \sum_{\ell = e,\mu,\tau} y^*_\ell \bar{N}^c_R S^{*} \ell_R  + \frac{1}{2}  \left(\bar{N}\right)_R^{c} M^{*}_{N} N_R $. In the case of more than one right handed neutrino, the second term has the same form with $M_N$ a symmetric matrix. Note that 
\begin{eqnarray*}
N_R = \frac{1}{2}\left(1 + \gamma_5\right) \Psi_D,  \left(N_R\right)^{c} =  \frac{1}{2}\left(1 - \gamma_5\right) \Psi_D
\end{eqnarray*}
where $\Psi_D$ is a 4 components spinor (i.e. Dirac spinor; which one could denote by just $N$).
} 
\begin{eqnarray}\label{eq:int:1}
\mathcal{L}_\textrm{Yuk} 
\supset \sum_{\ell=e,\mu,\tau} y_\ell \bar{\ell}^c_R S N_R +   \frac{1}{2} M_{N} \bar{N}_R \left(N_R\right)^{c}  + h.c.
\end{eqnarray}
The most general CP-invariant, and dimension four potential involving the SM Higgs doublet $\Phi$ and the charged Higgs scalar $S$ can be written as 
\begin{eqnarray}
V(\Phi, S) &=& - m_{11}^2 |\Phi^\dagger \Phi| + m_{22}^2 |S^\dagger S| + \lambda_1 |\Phi^\dagger \Phi|^2  + \lambda_2 |S^\dagger S|^2 + \lambda_3 |\Phi^\dagger \Phi| |S^\dagger S|,
\end{eqnarray}
with $m_{11}^2, m_{22}^2, \lambda_{i,i=1,2,3}$ are real parameters,  and $\Phi$ is the SM Higgs doublet  parametrised as 
$$
\Phi = \left (\begin{array}{c}
\omega^+ \\
\frac{1}{\sqrt{2}}(v + H + i \omega^0) \\
\end{array} \right).
$$ 
After the electroweak symmetry breaking, the pseudo-Goldstone bosons ($\omega^{0,\pm}$) get absorbed by the $Z/W^\pm$ gauge bosons, and one lefts with three new scalars; a CP-even scalar $H$ which is identified with the recently discovered $125$ GeV SM Higgs boson and a pair of charged scalars $H^\pm$. 
Their tree-level masses are given by: 
$$
m_H^2 = \lambda_1 v^2, \quad \textrm{and } \quad m_{H^\pm}^2 = m_{22}^2 + \frac{1}{2} \lambda_3 v^2.
$$

The non-trivial transformation of the scalar singlet under $U(1)_Y$ allows for gauge interactions with the photon and $Z$-boson. The gauge interaction of the $S$ state is given by
\begin{eqnarray}
\mathcal{L}_{\mathrm{gauge}} &=& - i ( e A^\mu - e \tan\theta_W Z^\mu) [S^\dagger (\partial_\mu S) - (\partial_\mu S^\dagger) S] \nonumber \\
&+& e^2 A_\mu A^\mu S^\dagger S + e^2 \tan^2\theta_W Z_\mu Z^\mu S^\dagger S - 2 e^2 \tan\theta_W A_\mu Z^\mu S^\dagger S, 
\end{eqnarray}
In addition to the SM parameters, the model includes the following  seven independent parameters
\begin{eqnarray}
\{m_{H^\pm}, m_N, \lambda_2, \lambda_3, y_e, y_\mu, y_\tau \}
\end{eqnarray}
For convenience, we define $y_N$ as the combination of the new Yukawa couplings as:
\begin{eqnarray}
 y_N = \sqrt{y_e^2 + y_\mu^2 + y_\tau^2}
\end{eqnarray}

\section{Theoretical and experimental constraints}
\label{sec:constraints}

The parameters of the model are subject to various theoretical and experimental constraints. Due to the fact that the charged Higgs state is gauge singlet, constraints from direct LHC searches are mainly coming from dilepton plus missing energy searches that  we will discuss in the next section. 
 Besides, the SM Higgs boson is really SM-like, i.e. there is no modification of tree-level Higgs couplings to fermions and gauge bosons. Therefore, the only modification to the Higgs boson decay rates
comes from the effect of the charged Higgs boson on the one-loop induced $H\to \gamma\gamma$ decay width \cite{Swiezewska:2012eh, Arhrib:2012ia, Jueid:2020rek}. In this work, we use the \textsc{Atlas}-\textsc{Cms} combined measurement of $|\kappa_\gamma| = \sqrt{\Gamma(H\to\gamma\gamma)/\Gamma(H\to\gamma\gamma)_\mathrm{SM}} = 0.87^{+0.14}_{-0.09}$  \cite{Khachatryan:2016vau}  in which no additional BSM  contributions are considered  to the Higgs boson decays except the one of the charged Higgs boson.\\

\begin{figure}[!t]
    \centering
    \includegraphics[width=0.48\linewidth]{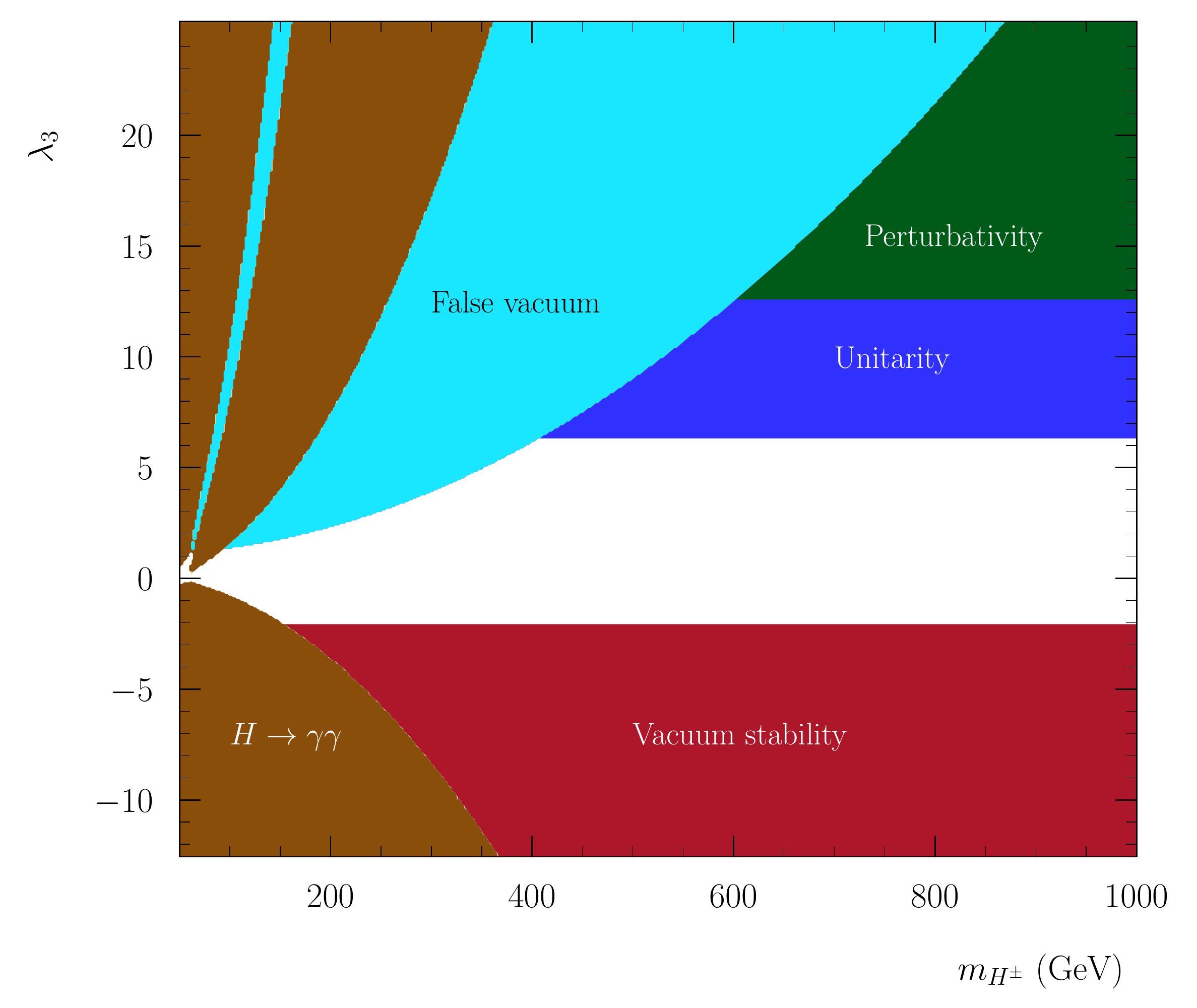}
    \hfill
    \includegraphics[width=0.495\linewidth, height=0.43\linewidth]{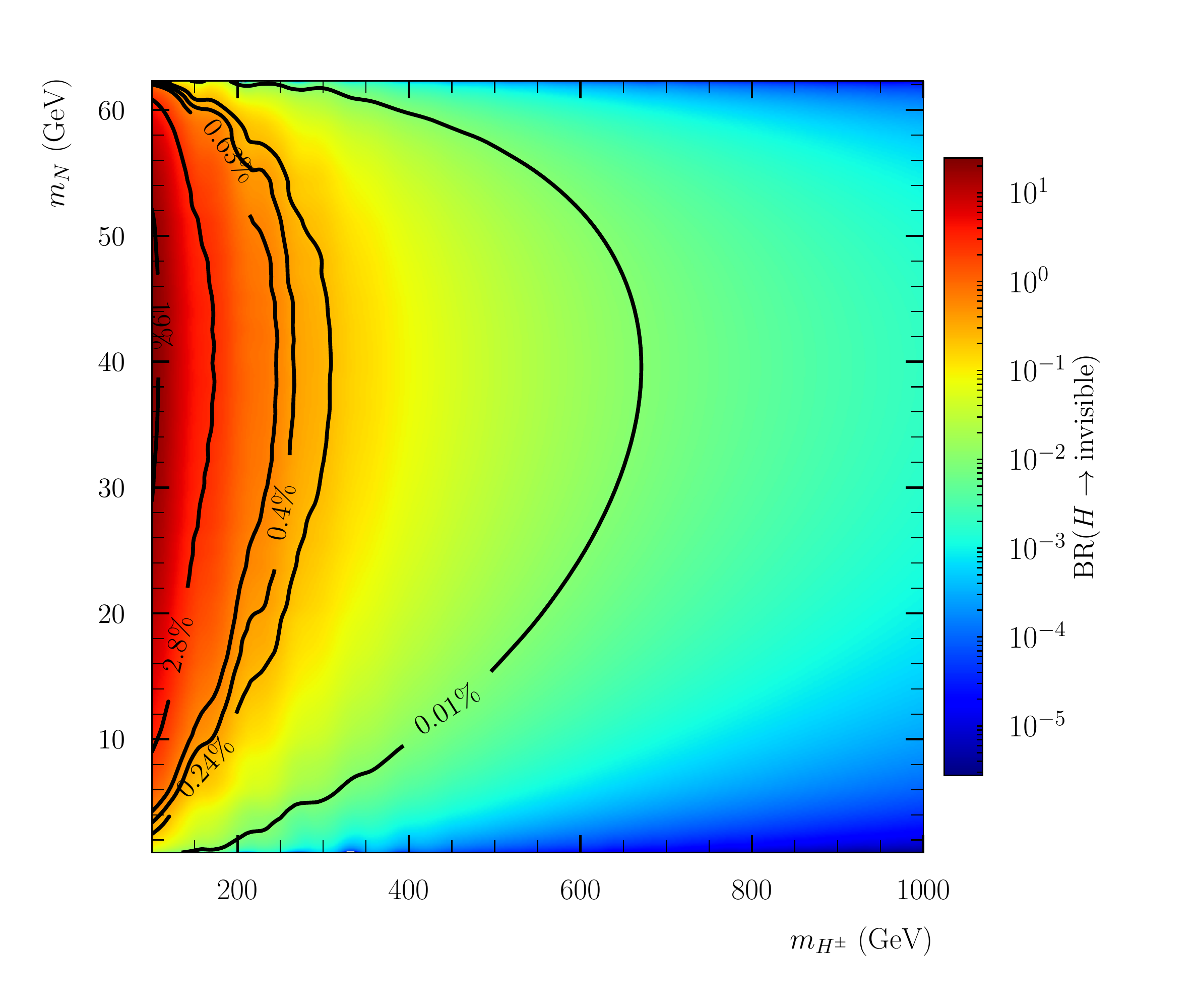}
    \caption{\emph{Left:} Summary of the impact of theoretical and experimental constraints on the parameter space of the scalar sector of the model. The constraints are depicted in the mass of the charged Higgs boson $m_{H^\pm}$ and $\lambda_3$ for $\lambda_2 = 2$.  The excluded regions are shown for constraints from $H\to \gamma\gamma$ (brown), from false vacuum (cyan), from vacuum stability (red), from perturbativity (green), and from perturbative unitarity (blue). The white shaded area corresponds to the allowed region of the 2D parameter space. \emph{Right:} The Higgs boson invisible decay branching projected on the mass of the charged Higgs ($m_{H^\pm}$) and the mass of the Majorana DM ($m_N$). The contours in solid black show the expected limits from HL-LHC \cite{CMS-PAS-FTR-16-002}, FCC-ee \cite{Cerri:2016bew}, ILC \cite{Asner:2013psa}, CEPC \cite{CEPCStudyGroup:2018ghi} and FCC-hh \cite{Selvaggi:2018obt} and correspond to upper bounds on $\mathrm{BR}(H\to \mathrm{invisible})$ ranging from $2.8\%$ to $0.01\%$. }
    \label{fig:constraints}
\end{figure}

In addition to constraint from the signal strength measurement, the model should conform to a number of theoretical constraints such as\footnote{Note that the theoretical constraints on our model can be obtained from those in the Inert Higgs Doublet Model (IHDM) by simply setting $\lambda_4 = \lambda_5 = 0$ (see e.g. \cite{Arhrib:2015hoa})} \\
\begin{itemize}
\item The quartic couplings of the scalar potential should satisfy the vacuum stability conditions (or boundedness-from-below) \cite{Branco:2011iw}.  These conditions can be written as: $\lambda_{1, 2} > 0,~ \lambda_3+2\sqrt{\lambda_1 \lambda_2} > 0$.
\item The perturbative unitarity should be maintained in various scattering-processes at high-energy \cite{Kanemura:1993hm, Akeroyd:2000wc}. This requirement can be  translated to a set of conditions on the combinations of the quartic couplings, i.e. $\max\{e_1, e_2, e_3\} < 4 \pi$, where $e_i$ represents the different eigenvalues of the scattering matrices given by $e_1 = \lambda_3$ and $e_{2,3} =- 3 (\lambda_1 + \lambda_2) \pm \sqrt{9 (\lambda_1 + \lambda_2)^2 + 4 \lambda_3}$.
\item The electroweak vacuum 
should be a global minimum \cite{Ginzburg:2010wa} which can be satisfied for $$
\frac{m_{11}^{2}}{\sqrt{\lambda_{1}}}\geq-\frac{m_{22}^{2}}{\sqrt{\lambda_{2}}}
$$
\end{itemize}
A summary of the theoretical and experimental constraints are shown in the left panel of Fig. \ref{fig:constraints}. The combination of all these constraints restricts the allowed range of $\lambda_3$ and the charged Higgs boson mass; namely   $\lambda_3$  cannot be larger than $5$ for all the charged Higgs masses.
The constraints on the Yukawa coupling $y_N$ come mainly from the limit on  branching fraction of the Higgs boson invisible decay mode; $\mathrm{BR}(H\to~\mathrm{invisble}) \equiv \textrm{B}_{\mathrm{inv}}$. This decay  occurs at  the one-loop order, which we have computed  
 using \texttt{FeynArts}, \texttt{FormCalc} and \texttt{LoopTools} \cite{Hahn:2000jm, Hahn:2000kx}. The strongest and up-to-date stringent bound on $\textrm{B}_{\mathrm{inv}}$ was reported by the \textsc{Cms} collaboration using  a combination of  previous Higgs to invisible decay searches at 7,8 and 13 TeV, where it has been found $B_{\mathrm{inv}} < B_{\mathrm{upper}} = 0.19$ at $95\%$ CL \cite{Sirunyan:2018owy} assuming SM production rates of the Higgs boson. Using the results of these searches, we derive an upper bound on the magnitude of $y_N$
\begin{eqnarray}
y_N < \bigg(\frac{2048 \pi^5 \Gamma_{H}^{\mathrm{SM}}}{\beta_N^{3/2} m_H \lambda_3^2 v^2 m_N^2 |C_0 + C_2|^2 \left(\frac{1}{B_\mathrm{upper}} - 1\right)}\bigg)^{1/4},
\end{eqnarray}
with $\beta_N = (1 - 4 m_N^2/m_H^2)$, and $C_{0,2} \equiv C_{0,2}(m_N^2,m_H^2,m_N^2,m_\ell^2,m_{H^\pm}^2,m_{H^\pm}^2)$ are the Passarino-Veltman functions \cite{Passarino:1978jh}. For instance, for $m_N \simeq 40$ GeV, $\lambda_3 = 1$ and $m_{H^\pm} = 200$ GeV, we get $y_N < 2.5$. In the right panel of Fig. \ref{fig:constraints}, we illustrate the predicted values of the $\textrm{BR}(H\to~\mathrm{invisible})$ projected on the $(m_{H^\pm}, m_{N})$ plane for $\lambda_3 = 1$ and $y_N = 2$. In  the same panel, we show the corresponding contours of the expected bounds from HL-LHC \cite{CMS-PAS-FTR-16-002}, FCC-ee \cite{Cerri:2016bew}, ILC \cite{Asner:2013psa}, CEPC \cite{CEPCStudyGroup:2018ghi} and FCC-hh \cite{Selvaggi:2018obt}. We can see that the current LHC bounds do not play a significant role in restricting the 2D parameter space. However, the expected bounds from FCC-hh can exclude charged masses up to $600~$GeV for $m_N \in [40, 50]~$GeV.\\

\begin{figure}[!t]
    \centering
    \includegraphics[width=0.70\textwidth]{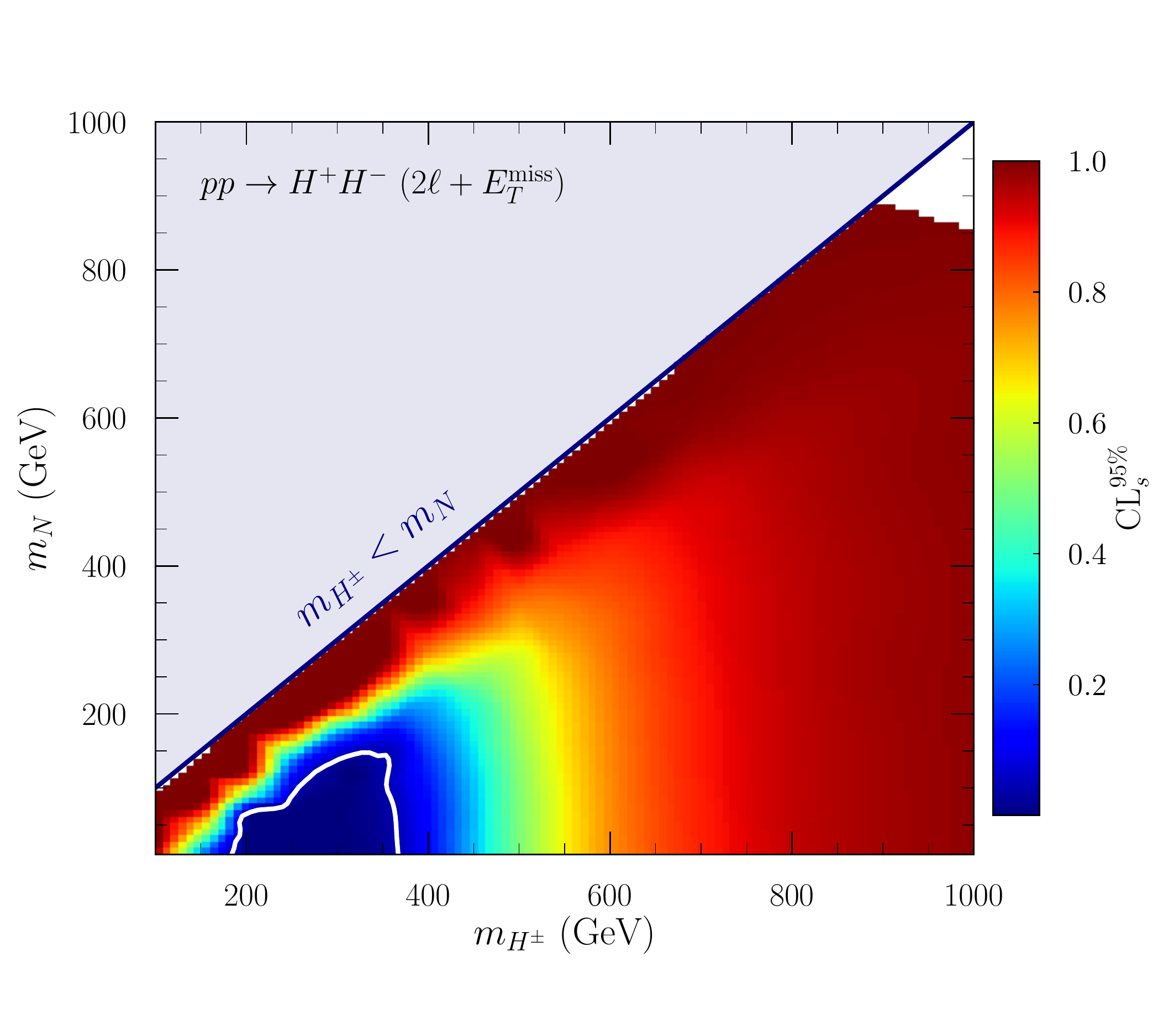}
    \vspace{-.9cm}
    \caption{The LHC exclusions from slepton searches at $\sqrt{s}=13~$TeV and $\mathcal{L} = 139~$fb$^{-1}$. The limits are projected on the mass of the charged Higgs and the mass of the Majorana DM. The white solid line corresponds to $\mathrm{CL}_s = 0.05$ below which the parameter point is excluded at the $95\%$ CL.}
    \label{fig:CLs_LHC}
\end{figure}

The experimental collaborations at LEP have carried out several searches of new physics scenarios such as the chargino and neutralino production at center-of-mass energies $\sqrt{s} = 183$-$209$ GeV. For instance, the OPAL collaboration has searched for new physics signal in $\ell^+ \ell^- + E_{T}^{\rm miss}$ final state using $680$ pb$^{-1}$ of luminosity \cite{Abbiendi:2003ji} whose null results can be used to put stringent bounds on the masses and couplings of charged scalars. In contrast, these obtained constraints do not affect significantly the model-parameter space; we found $m_{H^\pm} > 90$ GeV for $y_N \simeq 1$-$2$ (for more details, see section 2 in \cite{Ahriche:2018ger} and for further details on the IHDM analysis see \cite{Blinov:2015qva}). 

It is worth mentioning that constraints from lepton-flavor violating decays on the $y_\ell$ couplings are very important as it will be discussed in a later stage, where the most stringent bound comes from the branching ratio of $\mu \to e \gamma$ decay. Using the most recent bounds were used from the \textsc{Meg} \cite{Adam:2013mnn} and \textsc{BaBar} \cite{Aubert:2009ag} experiments,  we derive the following bounds: 
\begin{eqnarray}
|y_e y_\mu| < \bigg(\frac{2.855 \times10^{-5}}{\mathrm{GeV}} \bigg)^2 \frac{m_{H^\pm}^2}{|\mathcal{F}(m_N^2/m_{H^\pm}^2)|}, \nonumber \\
|y_e y_\tau| < \bigg(\frac{4.428 \times10^{-4}}{\mathrm{GeV}} \bigg)^2 \frac{m_{H^\pm}^2}{|\mathcal{F}(m_N^2/m_{H^\pm}^2)|}, \nonumber \\
|y_\tau y_\mu| < \bigg(\frac{4.759 \times10^{-4}}{\mathrm{GeV}} \bigg)^2 \frac{m_{H^\pm}^2}{|\mathcal{F}(m_N^2/m_{H^\pm}^2)|},
\end{eqnarray}
Here $\mathcal{F}(X)$ is the loop-function which can be found in \cite{Toma:2013zsa, Ahriche:2017iar}. The implications of these constraints on the product of the couplings are very important. For example, if we fix $y_e = 2$, we found that $\{y_e y_\mu, y_\tau y_e, y_\tau y_\mu \} < \{1.5$-$6.7 \times 10^{-4}, 0.05$-$0.1, 0.20$-$1.82 \times 10^{-5} \}$\footnote{Different scenarios where we can have $y_e \simeq y_\mu \simeq y_\tau \simeq 10^{-3}$ or $y_\tau \simeq y_\mu \gg y_e \simeq 10^{-4}$  are possible as well,  but none of these scenarios are phenomenologically plausible since the production rates at lepton colliders will be extremely small -- the cross sections depend exclusively on $y_e$. Theoretically, it could be possible to have a discrete flavor symmetry which allows for $y_e \simeq \mathcal{O}(1) \gg y_\tau \gg y_\mu.$}. In the following section, we consider an electron-specific scenario  where we chose  $y_e = 2$, $y_\mu = 10^{-4}$ and $y_\tau = 10^{-1}$.

\section{LHC constraints}
\label{sec:LHC}
The Yukawa interaction in \ref{eq:int:1}  is  subject to constraints from LHC searches of sleptons. This is due to 
the fact that for such interaction, the charged Higgs boson can be pair produced
 through $q\bar{q}$ annihilation with the exchange of $\gamma^*/Z$-bosons and its decay products lead to 2 $\ell + E_{T}^{\rm miss}$ final state. 
 In order to explore the severe restricting power of the LHC, we  take into account the most recent search for electroweak production of supersymmetric particles (sleptons/charginos)  by the ATLAS collaboration at $\sqrt{s} = 13~$TeV  with a recorded luminosity  $\mathcal{L} = 139~$fb$^{-1}$ \cite{Aad:2019vnb}.  For this purpose, exclusive samples of charged Higgs boson pair production have been generated at the Lowest Order (LO) with up to extra partons in the final state using \texttt{Madgraph\_aMC@nlo} \cite{Alwall:2014hca}. The MC signal samples were generated using a \texttt{UFO} format model file \cite{Degrande:2011ua}. The hard-scattering parton-level matrix elements are convoluted with the \texttt{NNPDF3.0lo} \cite{Ball:2014uwa} with $\alpha_s(M_Z^2) = 0.1118$ using the \texttt{LHAPDF} library \cite{Buckley:2014ana}, and the signal events were decayed using \texttt{MadSpin} \cite{Artoisenet:2012st}. The merging of the exclusive samples were performed using the \texttt{MLM} merging scheme \cite{Mangano:2006rw}  in addition to \texttt{Pythia8} \cite{Sjostrand:2014zea} using a merging scale that it is equal to the quarter of the charged higgs boson mass. Moreover, jets are clustered 
using \texttt{FastJet} \cite{Cacciari:2011ma} with the anti-$k_T$ algorithm with the separation parameter $R=0.4$ \cite{Cacciari:2008gp}.  The merged cross section values vary from $\sigma_{pp\to H^+H^- jj} = 81.12~\mathrm{fb}$ for $m_{H^\pm} = 100~$GeV to $\sigma_{pp\to H^+H^- jj} = 0.17~\mathrm{fb}$ for $m_{H^\pm} = 500~$GeV. 
Our results were found to be in excellent agreement with the slepton pair production at the ATLAS collaboration that reported  $\sigma_{pp\to \tilde{\ell} \tilde{\ell}} = 0.18\pm0.01~$fb for $m_{\tilde{\ell}} = 500~$GeV. \\
The ATLAS search strategy is based on selecting events that have exactly two oppositely charged signal leptons, with $p_T > 25$ GeV. All jets are predefined with $p_T > 20~$GeV and $|\eta| < 2.4$. Furthermore, events are then considered if the lepton pairs invariant mass $m_{\ell\ell} > 100$ GeV. 
The passed events have to fulfill another additional requirement  where no reconstructed $b$-jets are selected in each event.
These events  satisfy extra requirements where $E_{T}^{\mathrm{miss}} > 110$~GeV and MET significance $> 10$. Additional constraint is explored by studying the transverse mass $M_{T2}$ \cite{Lester:1999tx, Barr:2003rg},  defined as
\begin{eqnarray}
M_{T2} = \min_{q_{T}^{(1)} + q_{T}^{(2)} = {p}_{T}^{\mathrm{miss}}}\bigg\{\max\{M_T(p_{T}^{(1)}, q_{T}^{(1)}), M_T(p_{T}^{(2)}, q_{T}^{(2)})\} \bigg\},
\end{eqnarray} 
where $p_{T}^{(1)}$ and $p_{T}^{(2)}$ are the three-momentum of the leading and subleading charged leptons, respectively;  whereas ${q}_{T}^{(1)}$ and $q_{T}^{(2)}$ represent the three-momentum vectors of the unknown particles which can be combined in one vector $q_{T}^{(1)} + q_{T}^{(2)} = {p}_{T}^{\mathrm{miss}}$. The transverse mass in this case is given by $M_T(p_{T}^{i},q_{T}^{i}) = \sqrt{2 \times |p_{T}^{i}| \times |q_{T}^{i}| \times (1 - \cos\Delta\phi)}$. The $M_{T2}$ kinematical variable  can be used to fasten
the masses of the particles pair that can possibly escape the detection in each selected event. Hence, this will lead to reliable discrimination of the signal regions (SR).
Depending on the topology of the events, they will be   divided into two categories: same flavor (SF) events (where $m_{\ell \ell} > 121.2~$GeV cut is imposed) and different Flavor (DF) events. Moreover, an extra event classification is taken into account based on the multiplicity of non-$b$-tagged jets ($n_{\mathrm{j}}$) in each events where its content could have $n_{\mathrm{j}} = 0$ 
or $n_{\mathrm{j}} = 1$.
In our model, the charged Higgs decays branching ratios are: $\mathrm{BR}_{H^\pm \to e^\pm N} = 99.76\%$, $\mathrm{BR}_{H^\pm \to \tau^\pm N} = 2.33 \times 10^{-3}$, and $\mathrm{BR}_{H^\pm \to \mu^\pm N} \simeq 3.58 \times 10^{-9}$. Therefore, the strongest limits come from the signal region with SF leptons and $n_{\mathrm{j}} = 0$ or $n_{\mathrm{j}} = 1$, where these regions are electron enriched. For this purpose, the \texttt{MadAnalysis} 5 framework \cite{Conte:2012fm,Conte:2014zja,Conte:2018vmg, Fuks:2021wpe, Araz:2020dlf} is used to obtain the provided constraints from the LHC searches. The CL$_s$ value for each point in the parameter space is computed to obtain the exclusion bound at the $95\%$ confidence level \cite{Read:2002hq} (more details can be found in appendix \ref{sec:MA5}), and in Fig. \ref{fig:CLs_LHC} we give projected values on the ($m_{H^\pm}$, $m_N$) plane. The strongest limit comes from the 
signal region with SF, $n_{\mathrm{j}} = 0$ and $M_{T2} \in [160, \infty)$. Furthermore, charged Higgs masses up to $\simeq 400~$GeV can be excluded at $95\%$ CL considering the actual search (the additional limits from different signal regions are described in Appendix \ref{sec:MA5}).

\begin{figure}[!h]
\centering
\includegraphics[width=0.99\columnwidth]{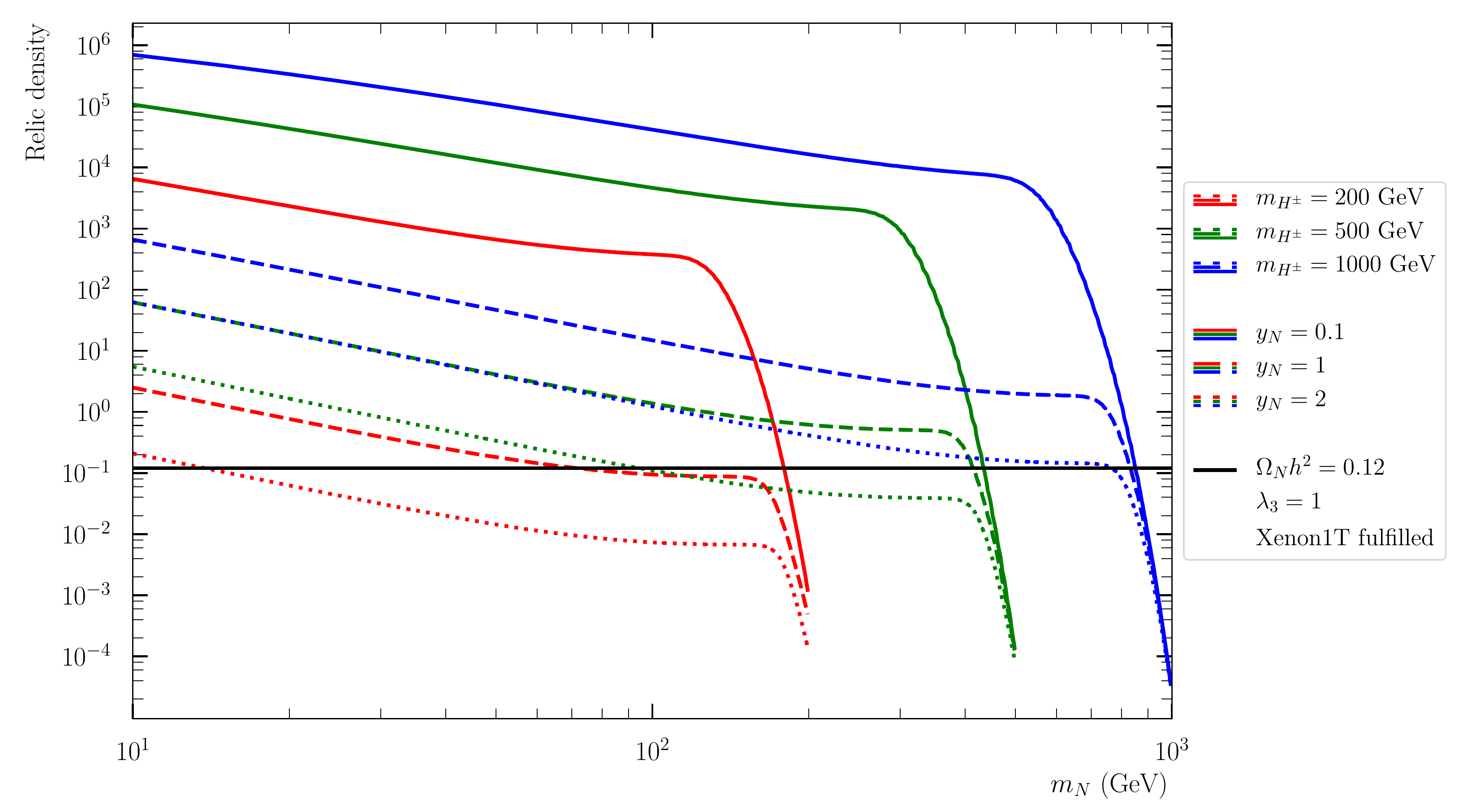}
\caption{Dependence of the DM  relic abundance as a function of  its  mass $m_N$ for various values of $y_N = \sqrt{y_e^2 + y_\mu^2 + y_\tau^2} \simeq y_e$ and charged Higgs mass $m_{H^\pm}$; the results are shown for $y_N = 0.1$ (solid), $y_N = 1$ (dashed), $y_N = 2$ (dotted). The color codings for the charged Higgs masses are as follows: red lines correspond to $m_{H^\pm} = 200$~GeV, green lines to $m_{H^\pm} = 500~$GeV and blue lines to $m_{H^\pm} = 1000~$GeV. All the values are consistent with bounds from the Xenon1T searches of DM-Nucleon elastic scattering.   All the calculations have been done for $\lambda_3 = 1$. The horizontal black solid line corresponds to the \textsc{Planck} measurement of $\Omega h^2 = 0.12$.}
\label{fig:DMrelic}
\end{figure}

\begin{figure}[!t]
    \centering
    \includegraphics[width=0.89\textwidth]{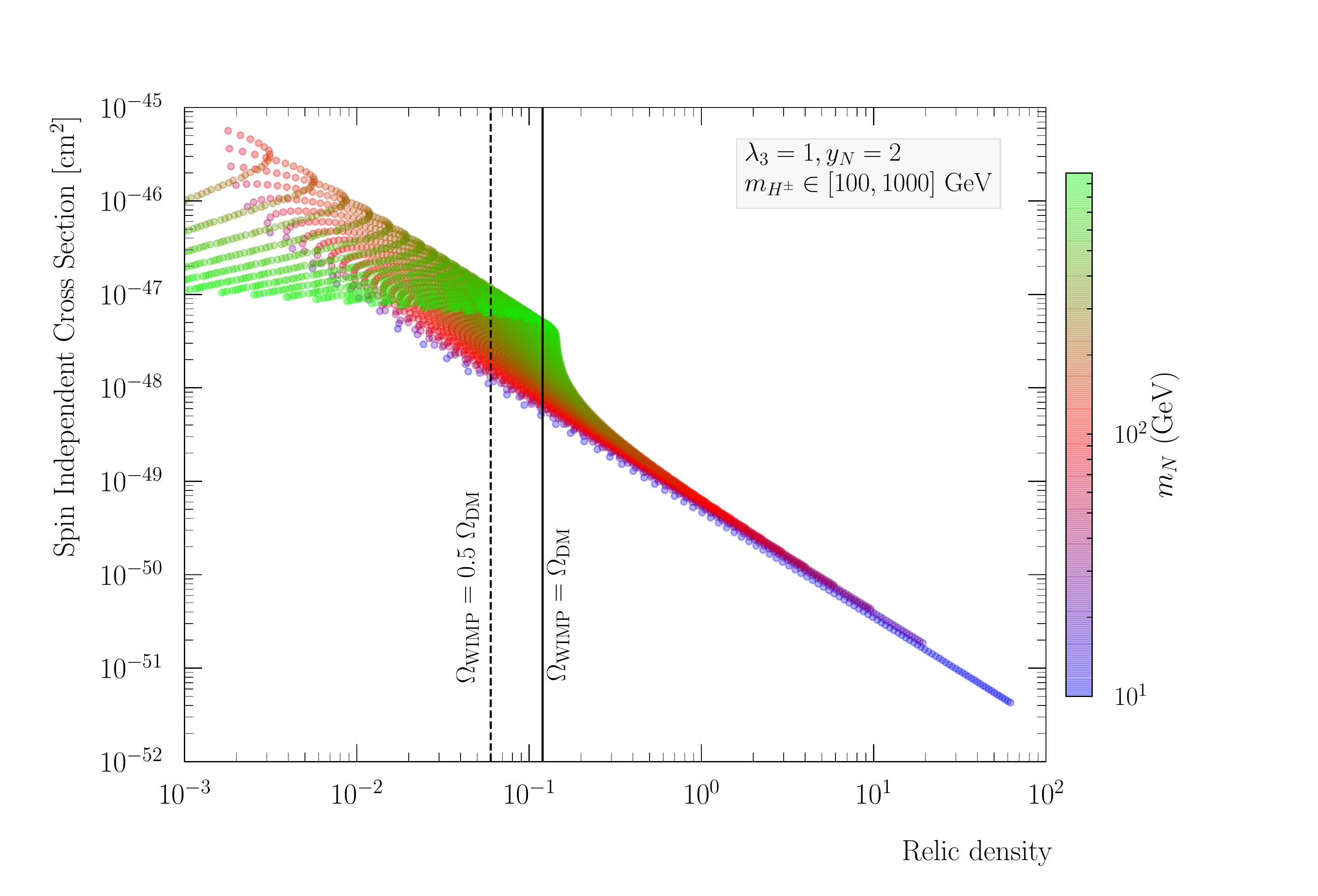}
    \caption{Correlation between the relic density and the spin-independent direct detection cross section. The palette shows the value of the DM mass. The two vertical lines correspond to $\Omega_N h^2 = \Omega_{\rm Planck} h^2$ (solid) and $\Omega_N h^2 = 0.5~\Omega_{\rm Planck} h^2$ (dashed).}
    \label{fig:Correlations}
\end{figure}

\section{Dark Matter Phenomenology}
\label{sec:DM}

The relic density of the $N$ particle can be approximated as follows \cite{Kong:2005hn}:
\begin{eqnarray}
 \Omega_N h^2 \simeq \frac{3 \times 10^{-26}~{\rm cm}^3 {\rm s}^{-1}}{\langle \sigma_{\rm eff}(x_f) \rangle}, 
\end{eqnarray}
where $\sigma_{\rm eff}(x_f)$ is the effective annihilation cross section at $x_f = m_N/T_f$  with  $T_f$ being  the freeze out temperature. The effective cross section is given by
\begin{eqnarray}
 \sigma_{\rm eff} (x_f) = \sum_{i,j}^2 \frac{4}{g_{\rm eff}^2} (1 + \Delta_i)^{3/2} (1 + \Delta_j)^{3/2} e^{-x_f (\Delta_i + \Delta_j)} \langle \sigma_{ij} v_r \rangle,
 \label{eq:sigEff}
\end{eqnarray}
In Eqn. \ref{eq:sigEff}, $g_{\rm eff}$ refers to the effective number of degrees of freedom $g_{\rm eff}(x_f) = \sum_{i=1}^2 2~(1+\Delta_i)^{3/2} e^{-x_f \Delta_i}$, $\sigma_{ij}$ refers to all the possible annihilation and co-annihilation cross sections and  $\Delta_i$ is the relative mass splitting defined by $\Delta_i = (M_i - m_N)/m_N~$($M_1=m_N,~ M_2=m_{H^\pm}$). In our analysis, we have included all possible annihilation and co-annihilation processes. For the annihilation, there are two major contributions: \emph{(i)} $NN\to \ell^+ \ell^-$ from the exchange of the charged Higgs boson in $t$- and $u$-channels, and \emph{(ii)} $NN \to \sum \bar{X} X$ (the sum is over all the SM particles) which arises from the exchange of the SM Higgs boson via the $s$-channel. In addition, the co-annihilation channels involve two generic contributions: \emph{(i)} $N H^\pm \to \ell^\pm H, W^\pm \nu, \ell^\pm Z, \ell^\pm \gamma$ and \emph{(ii)} $H^\pm H^\mp \to \ell^\pm \ell^\mp, q\bar{q}, HH, ZZ, W^\pm W^\mp, Z H, t\bar{t}$. The main contribution of the $N$ particles to the relic density comes from their annihilation  into charged leptons via the exchange of a charged scalar in the $t$- and $u$-channel. Moreover, the contribution of $s$-channel diagrams to the relic density is one-loop induced, and therefore is sub-leading which makes it below $1 \%$ of the total contribution. Note that  the annihilation rate in  the s-channel  does not depend significantly on $\lambda_3$ parameter if  the later had to satisfy the perturbativity bound. For small mass-splittings, i.e. $\Delta = (m_{H^\pm}-m_{N})/m_{N} < 0.1$, we find  that co-annihilations become very important and for some values of the model parameters (notably for $y_N < 1$)  dominates over the annihilation processes. \\

Now, we consider the scattering of the $N$ particles off a nuclei ($\mathcal{N}$), and since the  which 
occurs at the one-loop level  through the exchange of the SM Higgs boson. The corresponding spin-independent cross section is given by 
\begin{eqnarray}
 \sigma_{\rm SI} = \frac{4}{\pi} \mu_{\mathcal{N}}^2 \bigg( Z \cdot f_p + (A - Z) \cdot f_n \bigg)^2,
\end{eqnarray}
Here  $f_{p,n}$ are the nucleons ($p/n$) spin-independent form factors, and $\mu_{\mathcal{N}} \equiv m_N m_{\mathcal{N}}/(m_N + m_{\mathcal{N}})$ is the reduced mass of the DM-$\mathcal{N}$ system. The effective form-factors ($f_{p,n}$) are connected to the matrix element of the $q N$ scattering by: 
\begin{eqnarray}
 f_p = m_p \sum_{u,d,s} \frac{\mathcal{A}_q}{m_q} f^{Sq}_p + \frac{2}{27} m_p f_p^{Sg} \sum_{c,b,t} \frac{\mathcal{A}_q}{m_q}.
\end{eqnarray}
 For the proton, $\mathcal{A}_q$ is related to the leading-order matrix elements by
\begin{eqnarray}
 \mathcal{M}_{q N \to q N} = \mathcal{A}_q \langle p | \bar{\psi}_q \psi_q | p \rangle
\end{eqnarray}
where
\begin{eqnarray}
    \langle p | \bar{\psi}_q \psi_q | p \rangle = \bigg\{\begin{array}{lr}
        \frac{m_p}{m_q} \cdot f_{p}^{Sq}, & \text{for } q=u,d,s\\
        \frac{2}{27} \frac{m_p}{m_q} \cdot f_{p}^{Sg}, & \text{for } q=c,b,t,
        \end{array}
\end{eqnarray}
and
\begin{eqnarray}
 \mathcal{A}_q = \frac{\tilde{y}_{HNN}}{m_H^2} \cdot \frac{m_q}{v} \bar{\psi}_N \psi_N,
\end{eqnarray}
the effective $\tilde{y}_{HNN}$ is calculated at the one-loop order (see section \ref{sec:appendix1} for more details)
\begin{eqnarray}
\tilde{y}_{HNN} \simeq -\frac{\lambda_3 v |y_N|^2}{16 \pi m_N} \bigg[1 - \left(1 - r_N^{-2} \right) \log{\left(1 - r_N^2 \right)} \bigg], 
\end{eqnarray}
with $r_N = m_{N}/m_{H^\pm}$.
The effective low-energy nucleons form factors are obtained from chiral perturbation theory (please see \cite{DelNobile:2013sia, Hill:2014yxa, Bishara:2017pfq, Ellis:2018dmb} for a more detailed discussion). For the nuclear form factors the following values have been used: $f_p^{Su} = 1.53 \times 10^{-2}$, $f_p^{Sd} = 1.91 \times 10^{-2}, f_p^{Ss} = 4.47 \times 10^{-2},~$ and $f_p^{Sg} = 1 - \sum_{q=u,d,c} f_p^{Sq} = 0.9209$.
The spin-independent cross section scales as $\lambda_3^2 |y_N|^4/m_N^2$ modulo factors that depend on mass-splitting between the charged Higgs boson and the Majorana DM. To compute the relic abundance and Spin-independent Nucleus-DM elastic cross sections we use \texttt{MadDM} version 3.0 \cite{Backovic:2013dpa, Backovic:2015cra, Ambrogi:2018jqj}. In Fig. \ref{fig:DMrelic},  we show the relic density  as a function of the DM mass $m_N$ for various fixed mass values of the charged Higgs bosons and $y_N$. We can see that the relic abundance of the $N$ avoid the over-abundance bound from the \texttt{Planck} experiment if one of the following conditions are met: \emph{(i)} relatively large Yukawa coupling $y_N > 1$ and relatively light charged Higgs $m_{H^\pm} < 500~$ GeV and \emph{(ii)} for small mass-splittings, $\Delta < 0.1$ which is the regime dominated by co-annihilations regardless the possible values of $y_N$. In the remaining part of this paper, we consider the following benchmark scenarios\footnote{The collider implications of the scenario with $y_N < 0.1$ in which the DM relic abundance is dominated by the co-annihilation process will be studied in a future work.} 
\begin{equation}
100 \leqslant m_{H^\pm}/\mathrm{GeV} \leqslant 1000, \quad 10 \leqslant m_N/\mathrm{GeV} < m_{H^\pm},~\lambda_3 =1,~y_N = 2.
\end{equation}
A strong anti-correlation between $\Omega h^2$ and $\sigma_{\mathrm{SI}}$ can be seen in Fig. \ref{fig:Correlations}. The parameter space values produce a a DM relic density  $\Omega h^2 \simeq \Omega_\mathrm{Planck} h^2$ corresponding   masses in the range $ 200$-$750$ GeV for which the corresponding spin-independent cross section is about $\simeq 10^{-48} ~\mathrm{cm}^2$ which is below the \textsc{Xenon1T} bound \cite{Aprile:2018dbl}. Moreover,  as $m_{H^\pm}- N$ mass splitting  
gets smaller, the relic density decreases while still satisfying the \textsc{Xenon1T} bound. It is also worth noting that the model is unconstrained from the current indirect detection experiments due to the fact that $s$-channel annihilation channels are loop induced and mediated by the SM Higgs boson, with  the  predicted annihilation cross sections  $\langle \sigma v \rangle \simeq 10^{-37}~\mathrm{cm}^3/s$ that is orders of magnitude smaller than the  \textsc{Fermi-LAT} bounds \cite{Ackermann:2015lka}.

\begin{table}[!t]
\setlength\tabcolsep{4pt}
\begin{center}
\begin{tabular}{ c | ccc | ccc }
\toprule
$\sqrt{s}~[\mathrm{GeV}]$& \multicolumn{3}{c}{$500$} &
\multicolumn{3}{|c}{$1000$} \\ 
\hline
$[P_{e^-}, P_{e^+}]$ &   $[0,0]$   & $[+0.8, -0.3]$ & $[-0.8, +0.3]$ &  $[0, 0]$ & $[+0.8, -0.2]$ & $[-0.8, +0.2]$ \\
\hline
$W^+ W^-$ &   $7.19$ &   $1.06$ & $16.77$ &     $2.68$  & $0.44$ &  $5.79$      \\ 
$ZZ$   & $0.42$ & $0.32$ & $0.71$ &    $0.15$ &  $0.11$ &  $0.24$   \\ 
$t\bar{t}$   & $0.55$  &  $0.44$ &   $0.92$ &    $0.16$ &  $0.13$ &   $0.25$    \\ 
$H\nu\bar{\nu}_e$  & $7.76 \times 10^{-2}$  &  $1.14\times 10^{-2}$ & $0.18$ &  $0.21$ &  $3.44\times10^{-2}$ & $0.45$  \\ 
$HZ$  & $5.72\times 10^{-2}$  &  $5.72\times10^{-2}$ & $8.47\times10^{-2}$ &  $1.28\times10^{-2}$ &  $1.21\times10^{-2}$ & $1.77\times10^{-2}$  \\ 
\toprule
\end{tabular}
\end{center}
\caption{Production cross of the major background processes to the mono-Higgs channel in $e^+ e^-$ collisions at $\sqrt{s} = 500~$GeV and $\sqrt{s} = 1000~$GeV. The cross sections, given in picobarn, are calculated at the Leading-Order (LO) in perturbation theory.}
\label{table:ILC-MonoHiggs-Xsec}
\end{table}

\begin{table}[!h]
    \setlength\tabcolsep{9pt}
    \centering
    \begin{tabular}{c c c c c c}
    \toprule
    Cut     & $W^+W^-/ZZ$ & $t\bar{t}$ & $H\nu_e\bar{\nu}_e$ & $HZ$  \\
    \toprule
    Initial events  &   $2 \times 10^6~(100\%)$ & $10^6~(100\%)$ & $10^6~(100\%)$ & $10^6~(100\%)$ \\
    Lepton veto           &   $476982~(23.84\%)$ & $29449~(2.94\%)$ & $486875~(48.68\%)$ & $171353~(17.13\%)$ \\
    $n_b = 2$            &   $17885~(0.89\%)$  & $5275~(0.52\%)$ & $133594~(13.35\%)$ & $40031~(4.00\%)$  \\
    $p_T^{b\bar{b}} > 50~$GeV           &   $10025~(0.50\%)$ & $3572~(0.36\%)$ & $84701~(8.47\%)$ & $34975~(3.49\%)$  \\
    $E_{T}^{\rm miss} > 50~$GeV            &   $9559~(0.48\%)$ & $3175~(0.32\%)$ & $83846~(8.38\%)$ & $34286~(3.43\%)$ \\
    Signal region           &   $2~(2 \times 10^{-6})$ & $372~(0.037\%)$ & $18339~(1.83\%)$ & $733~(0.073\%)$ \\
    \toprule
    Initial events  &   $2 \times 10^6~(100\%)$ & $10^6~(100\%)$ & $10^6~(100\%)$ & $10^6~(100\%)$  \\
    Lepton veto           &   $432342~(21.61\%)$ & $15637~(1.56\%)$ & $493589~(49.35\%)$ & $203586~(20.35\%)$  \\
    $n_b = 2$            &   $15464~(0.77\%)$  & $3378~(0.34\%)$ & $138187~(13.82\%)$ & $48596~(4.85\%)$ \\
    $p_T^{b\bar{b}} > 50~$GeV           &   $9973~(0.49\%)$ & $2713~(0.27\%)$ & $102302~(10.23\%)$ & $42249~(4.22\%)$ \\
    $E_{T}^{\rm miss} > 50~$GeV            &   $8980~(0.45\%)$ & $2565~(0.25\%)$ & $101669~(10.16\%)$ & $42032~(4.20\%)$ \\
    Signal region           &   $1~(10^{-6})$  & $26~(0.0026\%)$ & $3541~(0.35\%)$ & $631~(0.063\%)$ \\
    \toprule
    \end{tabular}
    \caption{The cutflow table for various background contributions after each selection step. We show the cases  $\sqrt{s} = 500~$GeV (upper panels) and $\sqrt{s} = 1000~$GeV (lower panels). The number inside the parentheses show the efficiency after the selection step $i$ which is defined as $\epsilon = n_i/n_0$ with $n_0$ is the initial number of events.}
    \label{tab:efficiency}
\end{table}

\section{Prospects at the International Linear Collider}
\label{sec:ILC}

Our model gives rise to several signatures at the future International Linear Collider (ILC) that is expected to run at center-of-mass energies of $250, 350, 500$  and $1000$ GeV \cite{Baer:2013cma} with the setup of electron beams collisions \cite{Feng:1999zv,  Heusch:2005hb}. The advantage of the latter option is that it can be used efficiently  to probe lepton-violating  processes due to the small backgrounds contributions. At electron-positron colliders, there are several processes that can be used to probe the model; \emph{(i)}: $e^+ e^- \to N N Z$ which yields a signature dubbed mono-Z, \emph{(ii)}: $e^+ e^- \to N N \gamma$ which yields one photon in association with large missing energy,  \emph{(iii)}: $e^+ e^- \to N N H$ which yields a variety of final states depending on the decay products of the SM Higgs boson, \emph{(iv)}: $e^+ e^- \to H^\pm H^\mp$ which yields two opposite-sign charged leptons plus missing energy, and  \emph{(v)}: $e^- e^- \to H^- H^-$ which yields two same-sign charged leptons in addition to missing energy. We consider two signal processes; mono-Higgs process in $e^+ e^-$ collisions at $\sqrt{s} = 500,~1000$ GeV, and the same-sign charged Higgs pair production in $e^- e^-$ collisions at $1000$ GeV.\footnote{Other studies for DM production at lepton colliders have been performed in \cite{Bartels:2012ex, Andersen:2013rda, Ahriche:2014xra, Ko:2016xwd, Baouche:2017tzu, Kalinowski:2018kdn, Ghosh:2019rtj} in for different models.}

\begin{figure}[!t]
\centering
\includegraphics[width=0.495\linewidth]{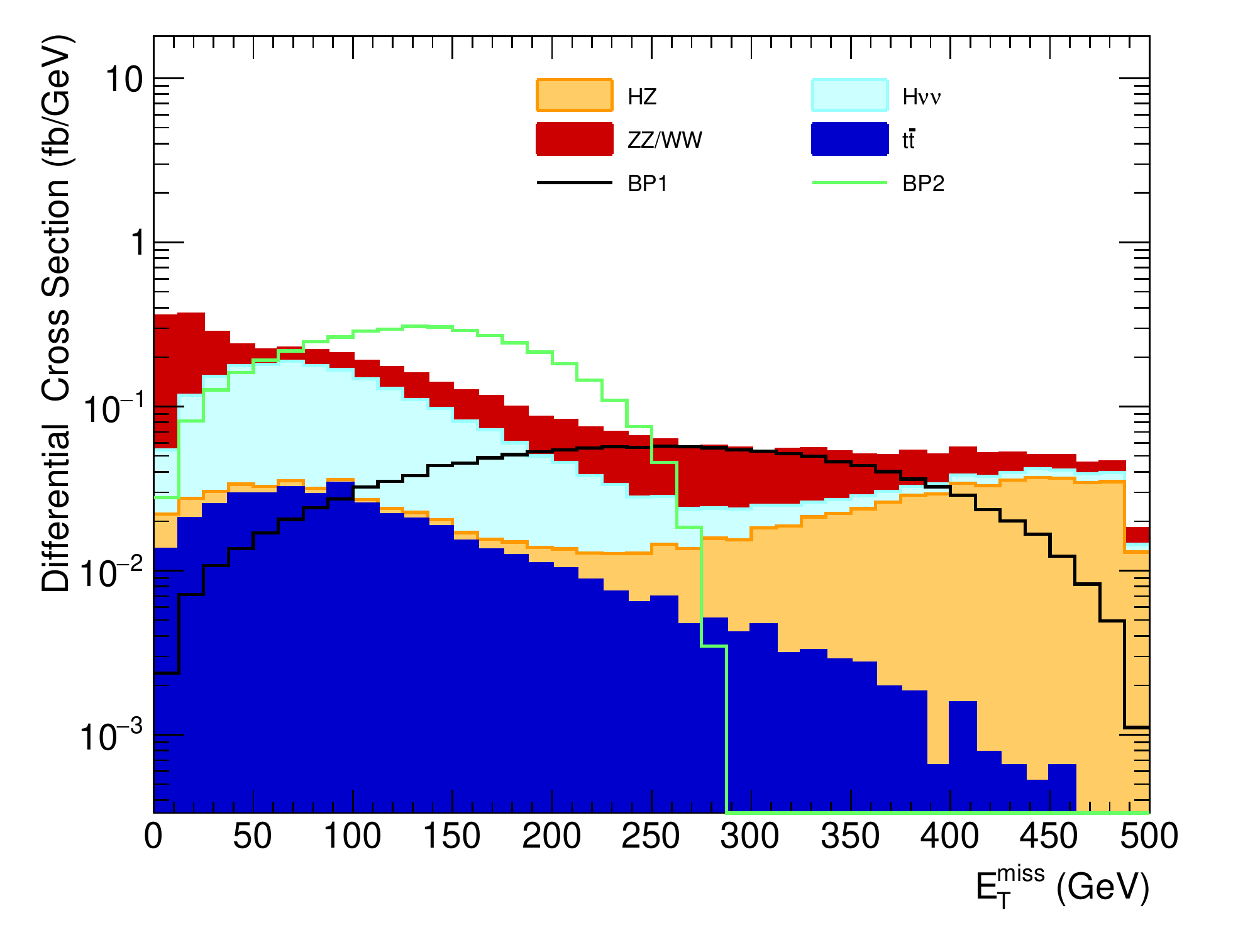}
\hfill
\includegraphics[width=0.495\linewidth]{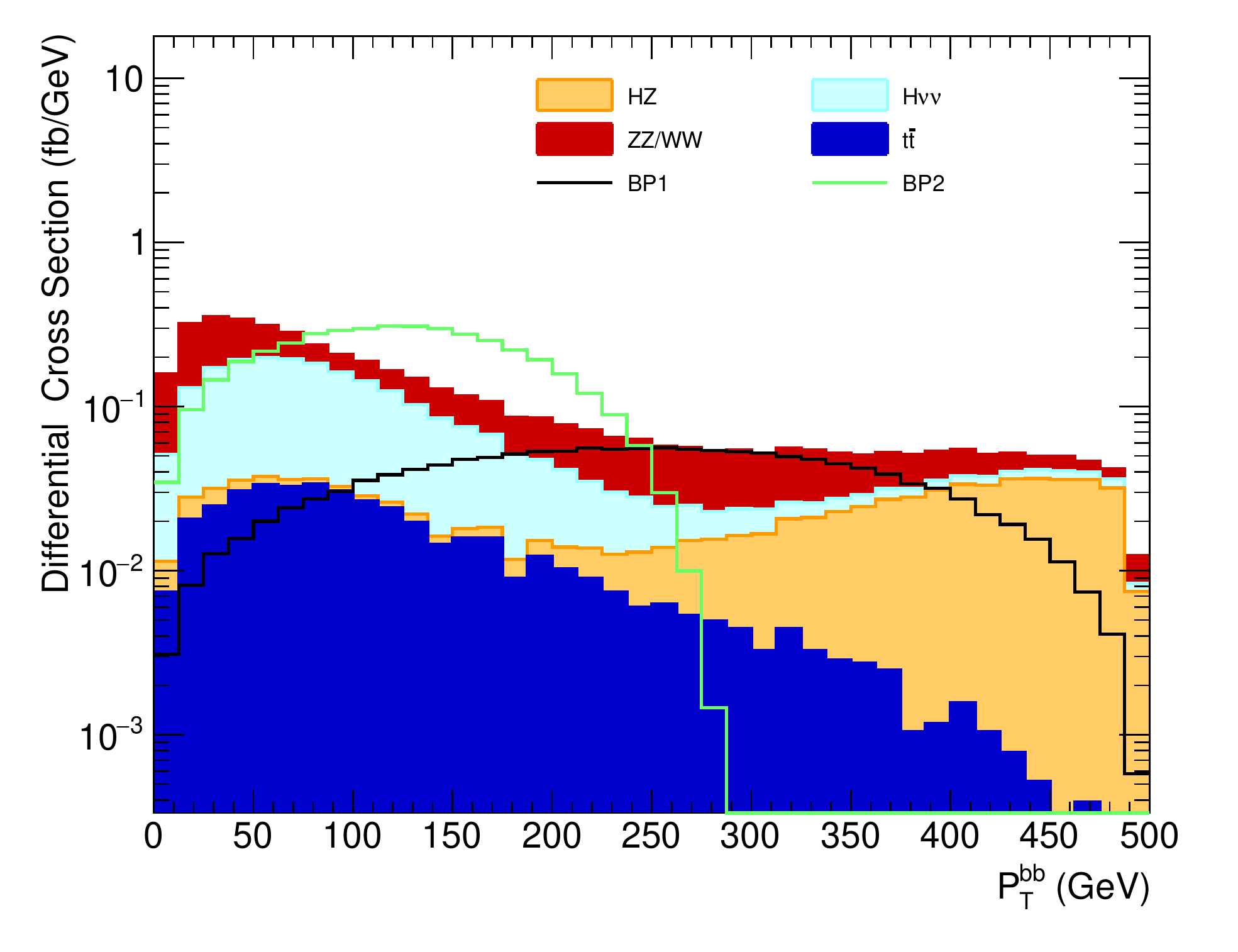}
\vfill
\includegraphics[width=0.495\linewidth]{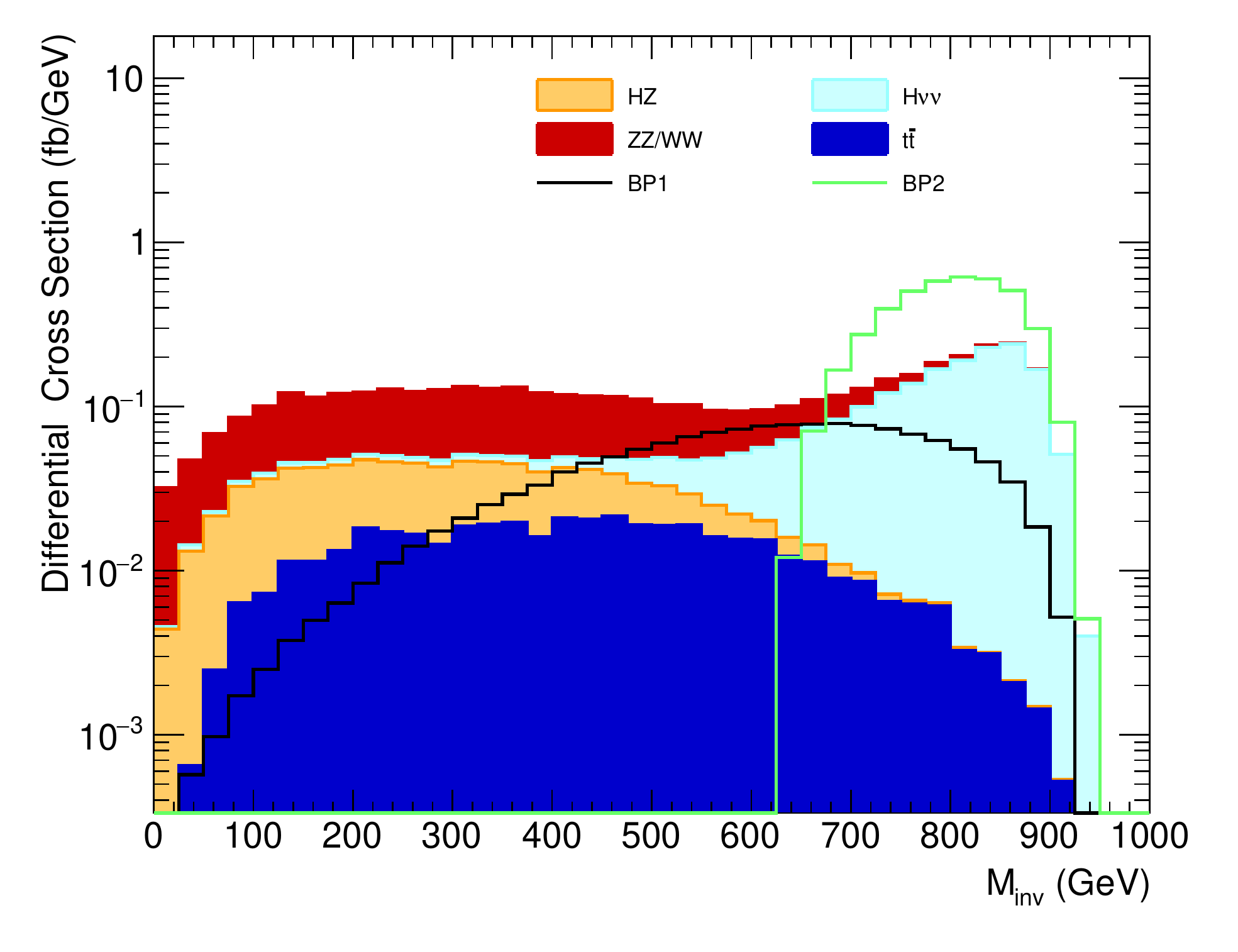}
\hfill
\includegraphics[width=0.495\linewidth]{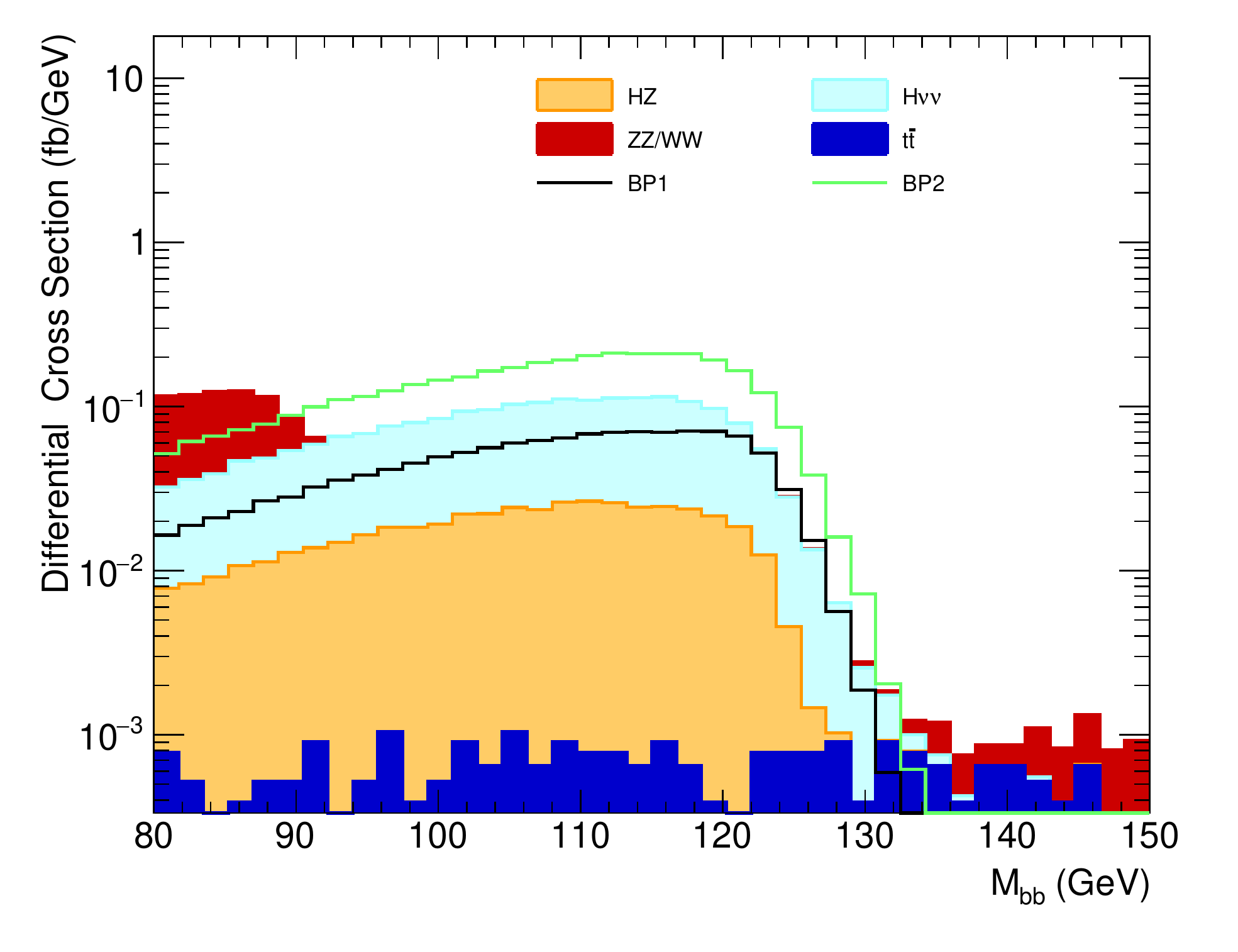}
\caption{Differential cross sections for some selected observables used in the signal-to-background discrimination in $e^+ e^- \to H + E_{T}^{\rm miss}$ process at $\sqrt{s} = 1000~$ GeV, and assuming $[P_{e^-}, P_{e^+}] = [+80\%, -20\%]$. From top left to bottom right, we display the missing transverse energy ($E_{T}^{\rm miss}$), the transverse momentum of the $b\bar{b}$ system ($p_{T}^{b\bar{b}}$), the invisible invariant mass ($M_{\rm inv}$), and the invariant mass of the $b\bar{b}$ system ($M_{b\bar{b}}$). We show the distributions for $HZ$ (orange), diboson ($ZZ/WW$), $H \nu_e \bar{\nu}_e$ (cyan) and $t\bar{t}$ (blue). For comparison, we show the differential cross sections for two signal benchmark points: \texttt{BP1} corresponding to $m_{H^\pm} = 1000~$GeV, and $m_N = 20~$GeV (black line) and \texttt{BP2} corresponding to $m_{H^\pm} = 460~$GeV, and $m_N = 320~$GeV (green line). The distributions are shown after the first two basic selections.}
\label{fig:ILC-MonoHiggs}
\end{figure}

For the mono-Higgs process, we consider the $H\to b\bar{b}$ decay channel of the SM Higgs boson for which the major backgrounds are $H(\to b\bar{b})Z(\to\nu\bar{\nu})$, $H(\to b\bar{b})\nu_e \bar{\nu}_e$, $W^+ W^-$, $ZZ$, and $t\bar{t}$ (the corresponding cross sections are given in Table \ref{table:ILC-MonoHiggs-Xsec}). Signal and backgrounds processes are generated using \texttt{Madgraph5\_aMC@NLO} \cite{Alwall:2014hca} and passed to \texttt{Pythia8} \cite{Sjostrand:2014zea} to add parton showering, hadronisation, and hadron decays. We employ  \texttt{Delphes} \cite{deFavereau:2013fsa} to account for detector  angularity, jet smearing, and particle resolutions. Jets are clustered according to the anti-$k_T$ algorithm \cite{Cacciari:2008gp} with $D=0.4$ using \texttt{FastJet} \cite{Cacciari:2011ma}. In Fig. \ref{fig:ILC-MonoHiggs}, we present the differential cross sections for some key variables that we used in the signal-to-backgrounds optimisation for signal process  and the main backgrounds. In that figure, we show the transverse momentum of the $b\bar{b}$ system ($p_T^{b\bar{b}}$), the missing transverse energy ($E_{T}^{\rm miss}$), the invariant mass of the $b\bar{b}$ system and of the reconstructed invisible momentum.\\

For the event selection, we start by considering  events that pass the following pre-selection criteria
\begin{itemize}
\item No lepton (electron or muon) with $p_T^\ell > 15$ GeV and $|\eta^\ell| < 2.5$.
\item Exactly two $b$-tagged jets with $p_T^b > 30$ GeV and $|\eta^b| < 2.5$.
\end{itemize}

Furthermore, the two $b$-tagged jets are used to form Higgs boson {\it candidate} whose transverse momentum is required to be larger than $50$ GeV.  The invariant mass of the invisible system is related to the energy of the Higgs boson candidate by the following relation 
\begin{eqnarray}
M_{\mathrm{inv}}^2 = s - 2 \sqrt{s} E_{b\bar{b}} + M_{b\bar{b}}^2,
\end{eqnarray} 
with $s$ being the center-of-mass energy, and  $E_{b\bar{b}}$ ($M_{b\bar{b}}$) is the energy (the invariant mass) of the $b\bar{b}$ system. We define the signal region by: \emph{(i)} the invariant mass of the $b\bar{b}$ system is required to satisfy $|M_{b\bar{b}} - m_H| <  10$ GeV, and \emph{(ii)} the invariant mass of the invisible system to satisfy $200~\mathrm{GeV}< M_{\mathrm{inv}} < 800~$GeV. For completeness, we show the cutflow for the different background contributions in Table \ref{tab:efficiency}. We note that the selection efficiency for both the signal as well as the backgrounds do not depend drastically on the polarisation setup of the incoming $e^+/e^-$ beams, and the the efficiency of the signal process varies in the range $2\%$-$3\%$. \\

\begin{figure}[!t]
    \centering
    \includegraphics[width=0.495\linewidth]{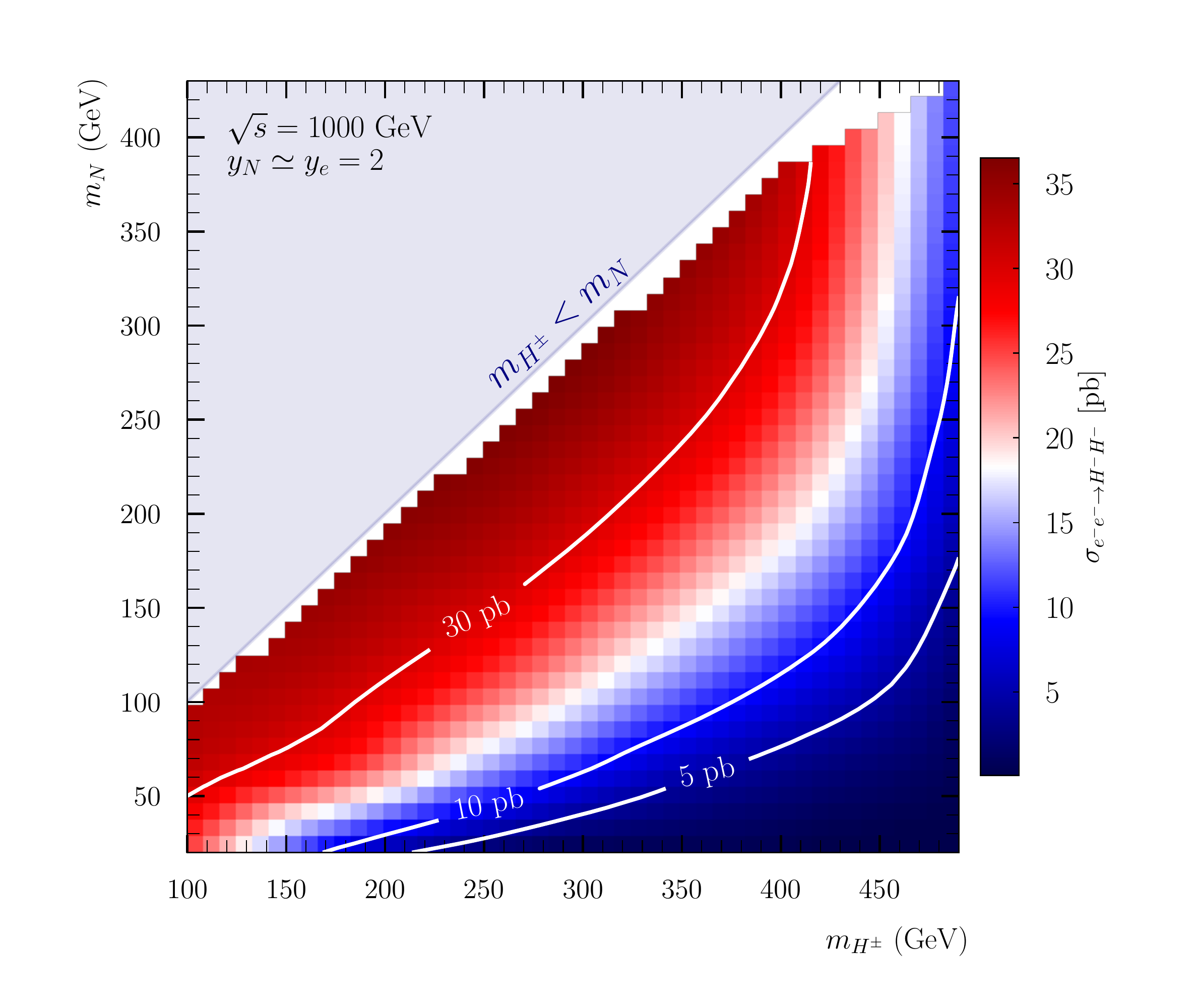}
    \hfill
    \includegraphics[width=0.495\linewidth]{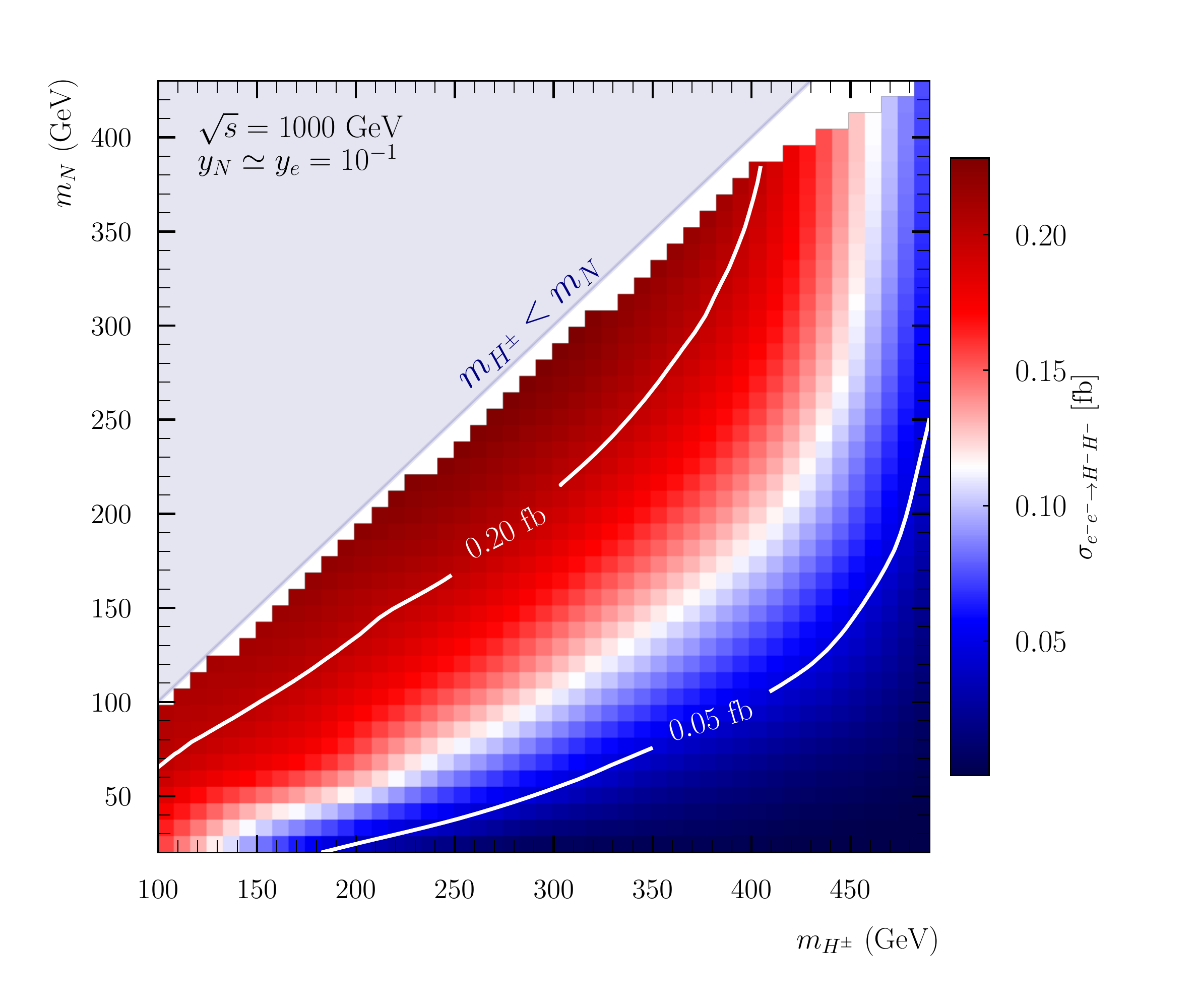}
    \caption{Production cross sections for same-sign charged Higgs pair production in $e^- e^-$ collisions at $\sqrt{s} = 1000~$GeV for $y_N \simeq y_e = 2$ (left panel) and $y_N \simeq y_e = 10^{-1}$ (right panel).}
    \label{fig:Xsec-Charged}
\end{figure}

The charged Higgs pair production in the electron-electron option at the ILC is certainly very interesting. The reasons are two-fold; \emph{(i)} the signal is enhanced for large DM masses due to the Majorana nature of the DM, and \emph{(ii)} the background is extremely suppressed since the SM processes conserve lepton number. Moreover, contrary to the mono-Higgs process, charged Higgs pair production can probe the whole parameter space of the model that can allowed by the collider energy. The cross section for the signal behaves as \cite{Aoki:2010tf}
\begin{eqnarray}
\sigma_{e^- e^- \to H^- H^-} \propto m_N^2 y_e^4,
\end{eqnarray}
Considering the decay of the charged Higgs boson into $e^- N$, which is the dominant one due to our choice of $y_\ell$ in section \ref{sec:constraints}, the main  backgrounds are $W^- W^- \nu_e \nu_e$ and $Ze^- e^-$ whose corresponding cross sections are $21.53$ fb and $98.1$ fb respectively. In Fig. \ref{fig:Xsec-Charged}, we display the cross section for the signal process for $y_N = 2$ (left panel). For comparison, we show in the right panel of Fig. \ref{fig:Xsec-Charged}, the cross section for $y_N = 0.1$. We can see that for $y_N = 2$, the signal cross section is orders of magnitude higher than the background ones; for $m_{H^\pm} = 250$ GeV  we have $\sigma_{e^- e^- \to H^- H^-}  = 5.46~(32.94)$ pb for $m_N = 30~(180)$ GeV. Given the large signal-to-background ratios, we only apply some minor selections on the decay products of the charged Higgs; i.e. two same-sign electrons with $p_T^e > 20$ GeV, and $|\eta^e| < 2.5$.

\begin{figure}[!t]
    \centering
    \includegraphics[width=1.01\columnwidth]{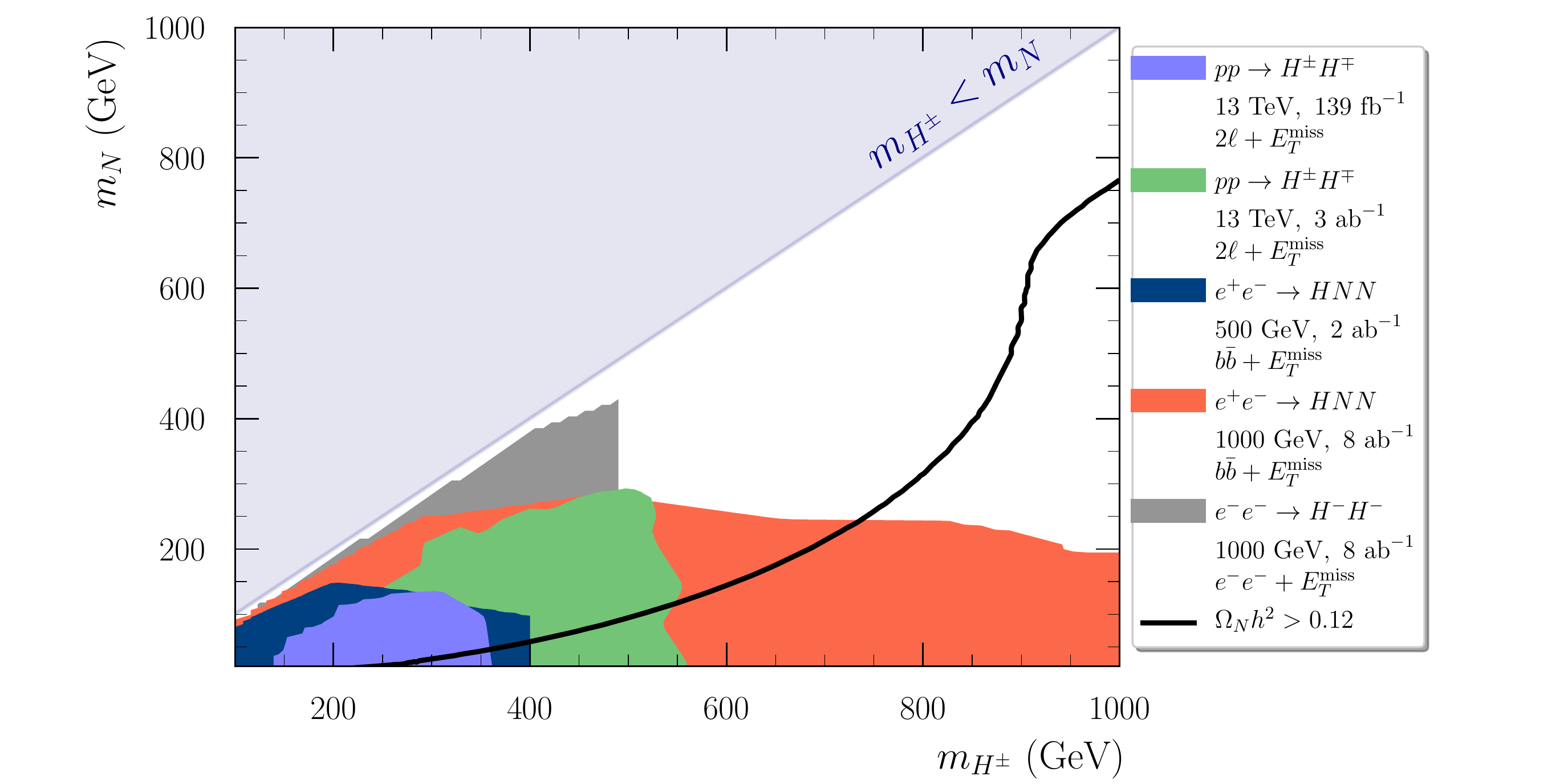}
    \caption{A summary of the impact of the present and future colliders on the parameter space of our model projected on the plane of the charged Higgs mass ($m_{H^\pm}$) and the mass of DM ($m_N$). The shaded purple area shows the excluded parameter space from the re-interpretation of sleptons searches at the LHC with $\mathcal{L} = 139~{\rm fb}^{-1}$. The green shaded region shows the extrapolations of the LHC limits to high luminosities ($\mathcal{L} = 3000~{\rm fb}^{-1}$). The expected exclusions from mono-Higgs searches at the ILC at both $\sqrt{s} = 500~{\rm GeV}$ and $\sqrt{s} = 1000~{\rm fb}^{-1}$ are shown in blue and red shaded areas respectively. Finally, the expected limits from searches of same-sign charged Higgs pair production at $\sqrt{s} = 1000~{\rm GeV}$ are shown as gray contour. We show the contour satisfying $\Omega_N h^2 = 0.12$ as black line (the area below which is excluded). All these results were obtained assuming $y_N \simeq y_e = 2$ and $\lambda_3 = 1$. The bounds from \texttt{Xenon1T} experiment are fulfilled.}
    \label{fig:sensitivity-ILCLHC}
\end{figure}

To obtain the expected exclusions on the parameter space from mono-Higgs process and the same-sign charged Higgs pair production at the ILC, we compute the CL$_s$ following the lines of section \ref{sec:MA5}. Since the number of observed events are not known yet, we assume it is equal to the expected number of backgrounds obtained from MC simulations at LO. On the other hand, we assume a flat $10\%$ on the expected background events. A  point in the space paramater of the model is expected to be excluded at the $95\%$ CL if the corresponding CL$_s$ is smaller than $0.05$. In  Fig. \ref{fig:sensitivity-ILCLHC}, we show the future sensitivity from slepton searches obtained from extrapolation of the exisiting LHC limits (green), mono-Higgs process (blue and red) and same-sign charged Higgs production (gray) on the charged Higgs mass and the mass of DM for $y_N = 2$ and $\lambda_3 = 1$. The following observations can be made:
\begin{itemize}
    \item The processes at hadron colliders cannot probe the region of the parameter space with small mass-splittings; i.e. $m_{H^\pm} - m_N < 50~{\rm GeV}$. The reason is that the charged leptons produced from charged Higgs production are relatively soft and do not pass the threshold of the lepton selection in addition to the strong refinement from transverse missing energy criteria. On the other hand, the ATLAS search (which was used in obtaining thse limits) relies on the transverse mass ($M_{T2}$) which has usually an end-point at the mass of the decaying charged Higgs bosons. Therefore, the inclusive signal regions defined by ATLAS cannot probe light charged Higgs bosons in our scenario.
    \item The mono-Higgs process can be used to probe DM masses up to $200~{\rm GeV}$ indenpendtly of the charged Higgs boson mass. However, this process is strongly dependent on both $\lambda_3$ and $y_N$. Decreasing the values of those parameters will reduce the expected event yields and, therefore, weakens the limits. 
    \item The rate of the same-sign charged Higgs pair production grows quadratically with the mass of the Majorana DM, and the associated backgrounds are essentially negligible. Therefore, this process can probe all the regions of the parameter space allowed by the collider energy provided that $y_N$ is of order unity or larger. 
\end{itemize}

\section{Conclusions}
\label{sec:conclusion}
In summary, we studied an interesting scenario which may provide DM candidate and at the same time an explanation for the absence of DM signals in both direct detection experiments and at colliders. In this framework, we minimally extended the SM by two gauge singlets: a charged scalar and a Majorana singlet fermion. First, we studied the impact of different theoretical and experimental constraints on the model parameter space, and  showed that it  can be severely constrained by limits from lepton-flavor violating decays at present experiments which in turn  would suppress the magnitude of the couplings of the DM candidate to the visible sector. On the other hand, Higgs boson invisible decays, if observed in future experiments, would further restricts the range of the couplings for relatively light DM; $m_{\rm DM} < m_H/2$. Then, using simple correlations between the relic abundance and the spin-independent cross sections, we showed that DM masses consistent with the \textsc{Planck} observations are still allowed by the \textsc{Xenon1T} bound, and well above  the neutrino-floor. Besides, we found that this model can be probed at lepton colliders  using the mono-Higgs process and the same-sign charged Higgs pair production. This simple model provides a one-to-one correspondence between its parameters and the predicted values for physical observables.  For instance, the mono-Higgs signature of the model can be used to probe both $m_N$ and the product of $\lambda_3$ and $y_N$. An inference on these parameters from the measurement of the rate of this process will restrict the dependence of the predictions for direct detection experiments on them -- we note that $\sigma_{\rm SI}$ involves an additional dependence on $m_{H^\pm}$. The same-sign charged Higgs boson pair production is anti-correlated to the relic density and is proportional to $y_N^4 m_N^2$. Therefore, for the set of parameters $\{\lambda_3, y_N, m_{H^\pm}, m_N\}$ there are four corresponding observables, i.e. $\{\sigma_{e^+ e^- \to HNN}, \sigma_{e^- e^-\to H^- H^-}, \sigma_{\mathrm{SI}}, \Omega h^2\}$. Important steps are yet to be made regarding the probes of these scenarios using one of the many developed methods to study the characteristics of DM at colliders. One can also  obtain the interaction \eqref{eq:int:1} from a more UV complete model. For instance, this can be realized by embedding the SM into SU(5) gauge group  with the matter fields in $10$ and $\bar{5}$ representation and the charged singlet belongs to the $10_{H}$ representation, while the right handed neutrino belongs to the singlet representation, i.e.~$\mathcal{L}_{\text{int}}= g_{\alpha \beta} \overline{10}_\alpha \otimes 10_{H} \otimes 1_{N_\beta}\supset g_{\alpha \beta} \ell_{R\alpha}^T C N_\beta S^+$. 
Another possibility is the flipped-$SU(5)\times U(1)_X$ grand-unified theory, where the right handed lepton field is  singlet of $SU(5)$,  and the right handed  neutrino is a member of  the $10$ representation. In this case,  the interaction \eqref{eq:int:1} can be obtained from the following effective Lagrangian $\mathcal{L}_{\text{int}} = \frac{h_{\alpha \beta}}{\Lambda} \overline{10}_\alpha \otimes \bar{1}_\beta \otimes 10_{H} \otimes 1_{S} + h.c. \supset \frac{h_{\alpha\beta} \langle 10_H \rangle}{\Lambda} N^T C \ell_R S^-$. The embedding of our model into a grand-unified theory is certainly an interesting question to pursue which we report on for a future study \cite{Jueid:2020xx}. \\

\section*{Acknowledgments}
AJ would like to thank the CERN Theoretical Physics Department for its hospitality where a part of this work has been done. AJ would like to thank Jack Araz and Benjamin Fuks for providing the necessary \texttt{MadAnalysis} routines corresponding to the implementation of the ATLAS-SUSY-18-032 search which we used in this work to obtain the LHC constraints on the model parameter space. Most of the plots in this paper were made using \texttt{Matplotlib} \cite{Hunter:2007}. Numerical and Statistical calculations were made using \texttt{NumPy} and \texttt{SciPy} \cite{2020NumPy-Array, 2020SciPy-NMeth}. The work of AJ is supported by the National Research Foundation of Korea, Grant No. NRF-2019R1A2C1009419. 

\appendix
\section{Computation of the effective $\tilde{y}_{HNN}$ coupling}
\label{sec:appendix1}

In this section, we show the detailed calculation of the effective $\tilde{y}_{HNN}$ coupling. To do so, we consider the process
\begin{eqnarray}
H(q) \longleftrightarrow N(p_1) + \bar{N}(p_2).
\label{eqn:reaction}
\end{eqnarray}
where in eqn. \ref{eqn:reaction}, $p_1$ and $p_2$ are the four-momenta of the right handed Neutrinos satisfying the on-shell constraint 
$p_1^2=p_2^2=m_N^2$ and $q$ is the four-momentum of the Higgs boson.

\begin{figure}[!t]
\centering
\includegraphics[width=0.87\linewidth]{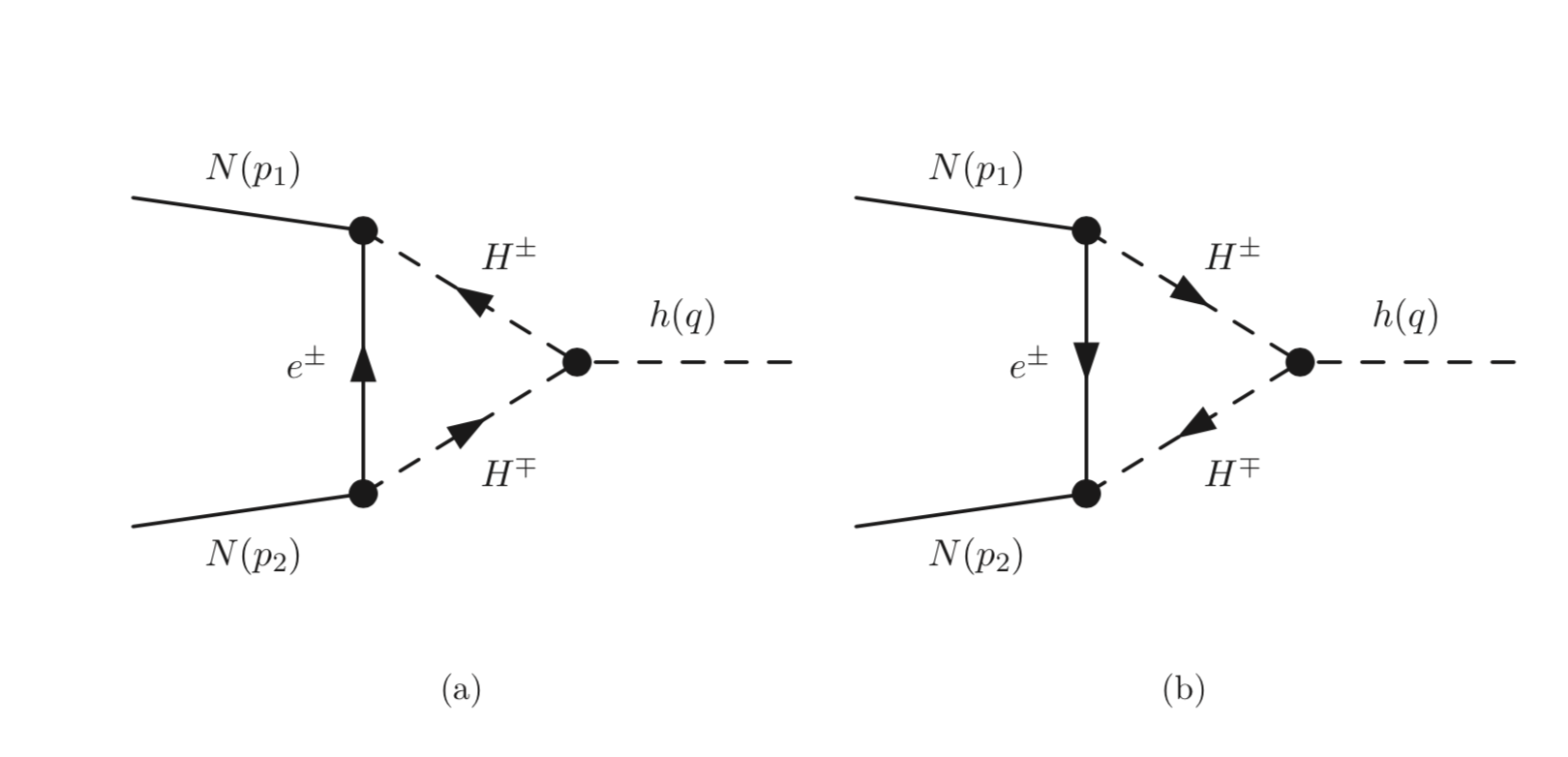}
\caption{One-loop Feynman diagram for the $H(q) \to N(p_1) N(p_2)$ process which gives rise to the effective $\tilde{y}_{HNN}$ coupling.}
\label{fig:hNNDiagram}
\end{figure}

The corresponding Feynman amplitude can be written as 
\begin{eqnarray}
\mathcal{M} &=& \sum_{\ell=e,\mu,\tau} \int \frac{\textrm{d}^4 k}{(2 \pi)^4} \frac{\bar{v}(p_2) (-iy_\ell^* P_R) i (\slashed{k} + m_\ell) (-iy_\ell P_L) (-i \lambda_3 v) u(p_1)}{(k^2-m_\ell^2)((p_2+k)^2 - m_{H^\pm}^2)((k-p_1)^2 - m_{H^\pm}^2)} + (R \longleftrightarrow L), \nonumber \\
&=& \sum_{\ell=e,\mu,\tau} \lambda_3 v |y_\ell|^2  \int \frac{\textrm{d}^4 k}{(2 \pi)^4} \frac{\bar{v}(p_2) P_R \slashed{k} P_L u(p_1)}{k^2((k+p_2)^2 - m_{H^\pm}^2)((k-p_1)^2 - m_{H^\pm}^2)} + (R \longleftrightarrow L).
\label{eqn:amp1}
\end{eqnarray}
In the second line, we have ignored the mass of the charged lepton in both the numerator and the denominator. After redefining the integration variables, and making few mathematical manipulations, we get:
\begin{eqnarray}
\mathcal{M} = \lambda_3 v |y_N|^2 \bigg(\bar{v}(p_2) P_R \mathcal{I}_{(I)}^\mu \gamma_\mu  P_L u(p_1) - \bar{v}(p_2) P_R \slashed{p}_2  \mathcal{I}_{(II)} P_L u(p_1)\bigg) + (\gamma_5 \to -\gamma_5),
\label{eqn:amp2}
\end{eqnarray}
where we have used $|y_N|^2 = \sum_{\ell=e,\mu,\tau} |y_\ell|^2$. The integrals we need to evaluate are 
\begin{eqnarray}
\mathcal{I}_{(I)} &=& \int \frac{\textrm{d}^4 k}{(2 \pi)^4} \frac{k^\mu}{(k-p_2)^2(k^2 - m_{H^\pm}^2)((k-p_1-p_2)^2 - m_{H^\pm}^2)} \nonumber \\
\mathcal{I}_{(II)} &=& \int \frac{\textrm{d}^4 k}{(2 \pi)^4} \frac{1}{k^2((k+p_2)^2 - m_{H^\pm}^2)((k-p_1)^2 - m_{H^\pm}^2)}.
\end{eqnarray}
Evaluating these integrals using dimensional regularization, one easily gets
 \begin{eqnarray}
\mathcal{I}_{(I)}^\mu &=& \frac{i}{16 \pi^2} \bigg(p_2^\mu C_2 + (p_1^\mu+p_2^\mu) C_1 \bigg), \quad \mathcal{I}_{(II)} = \frac{-i}{16 \pi^2} C_0.
\label{eqn:amp3}
\end{eqnarray}
with 
$$
C_i \equiv C_i(m_N^2, q^2, m_N^2, 0, m_{H^\pm}^2, m_{H^\pm}^2),~i=0,1,2.
$$ 
are the three-point Passarino-Veltman functions \cite{Passarino:1978jh}. Inserting the expressions of the integrals into eqn. \ref{eqn:amp2}, and using Dirac equations, we find:
\begin{eqnarray}
\mathcal{M} = \frac{\lambda_3 v |y_N|^2 m_N}{16 \pi^2} \bar{v}(p_2) u(p_1) \left(C_0 + C_2 \right).
\label{eq:ampFinal}
\end{eqnarray}
To get the effective $\tilde{y}_{HNN}$ coupling, we estimate the amplitude in equation \ref{eq:ampFinal} for $q^2 \simeq 0$. The Leading terms in the expansions of the Passarino-Veltman functions near $q^2 \simeq 0$ are given by\footnote{We have checked the correctness of our calculations by comparing the analytical expansions of the Passarino-Veltman function near $q^0 \simeq 0$ with the output we got using \texttt{Package-X} \cite{Patel:2015tea}.}
\begin{eqnarray*}
C_0 \simeq \frac{1}{m_N^2} \log\bigg(\frac{m_{H^\pm}^2 - m_N^2}{m_{H^\pm}^2}\bigg) + \mathcal{O}(q^2),  \\
C_2 \simeq \frac{-1}{m_N^4} \Bigg[m_N^2 + m_{H^\pm}^2 \log\bigg(\frac{m_{H^\pm}^2 - m_N^2}{m_{H^\pm}^2}\bigg) \Bigg] + \mathcal{O}(q^2).
\end{eqnarray*}
Therefore, we find the expression of the $\tilde{y}_{HNN}$ coupling at $q^2 \simeq 0$
\begin{eqnarray}
\tilde{y}_{HNN} = \frac{-\lambda_3 v |y_N|^2}{16 \pi^2 m_N^3} \Bigg[m_N^2 + (m_{H^\pm}^2 - m_N^2) \log\left(1 - \frac{m_N^2}{m_{H^\pm}^2}\right) \Bigg] + \mathcal{O}(q^2),
\end{eqnarray}
which is in excellent agreement with the results of \cite{Okada:2013rha, Ahriche:2017iar}. 

\begin{table}[!t]
\setlength\tabcolsep{11pt}
{\begin{tabular}{@{}ccccc@{}} \toprule \toprule
Region         & SR-SF-0J            & SR-SF-0J            & SR-SF-0J           & SR-SF-0J \\
$M_{T2}~$(GeV) & $\in [100,\infty)$  & $\in [160,\infty)$  & $\in [100,120)$    & $\in [120,160)$ \\
\toprule
Observed events & $147$              & $37$                & $53$               & $57$ \\
\toprule
Fitted backgrounds & $144\pm12$      & $37.3\pm3.0$        & $56\pm6$           & $51\pm5$ \\
\toprule
$m_{H^\pm}, m_{N} = (200, 10)~$GeV   & $109.6$ & $36.7$ & $23.9$  & $48.9$  \\
$m_{H^\pm}, m_{N} = (200, 142)~$GeV  & $1.7$ & $0.0$ & $1.5$ & $0.2$  \\
\toprule \toprule
Region         & SR-SF-1J            & SR-SF-1J            & SR-SF-1J           & SR-SF-1J \\
$M_{T2}~$(GeV) & $\in [100,\infty)$  & $\in [160,\infty)$  & $\in [100,120)$    & $\in [120,160)$ \\
\toprule
Observed events & $120$              & $29$                & $55$               & $36$ \\
\toprule
Fitted backgrounds & $124\pm12$      & $36\pm5$        & $48\pm8$           & $40\pm4$ \\
\toprule
$m_{H^\pm}, m_{N} = (200, 10)~$GeV & $53.7$ & $19.1$ & $11.3$ & $23.4$ \\
$m_{H^\pm}, m_{N} = (200, 142)~$GeV & $1.3$ & $0.0$  & $1.2$  &  $0.1$ \\
\toprule \toprule
\end{tabular}
\caption{Observed and expected background event yields for the `same flavor' (SF) inclusive signal regions SRs (those numbers are taken from \cite{Aad:2019vnb}). For reference, we show the number of events in the SRs for few benchmark points in our model corresponding to $(m_{H^\pm}, m_N) = (200, 10), ~\mathrm{and} (200, 142)~$GeV.}
 \label{tab:NeventsLHC}}
\end{table}

\section{LHC limits: Signal Regions (SR) and the statistical treatment}
\label{sec:MA5}
In this section, we describe some details about the statistical setup, the exclusion bounds for all the considered inclusive signal regions as well as the method used to obtain the limit extrapolation up to higher expected luminosities. The ATLAS collaboration has tackled such an  analysis by defining four inclusive signal regions for each jet bin category, based on the $M_{T2}$ interval splitting variable. The summary of the expected background and observed event numbers is given in Table \ref{tab:NeventsLHC} with different signal region definitions. \\

\begin{figure}[!tbp]
\centering
\includegraphics[width=0.85\linewidth]{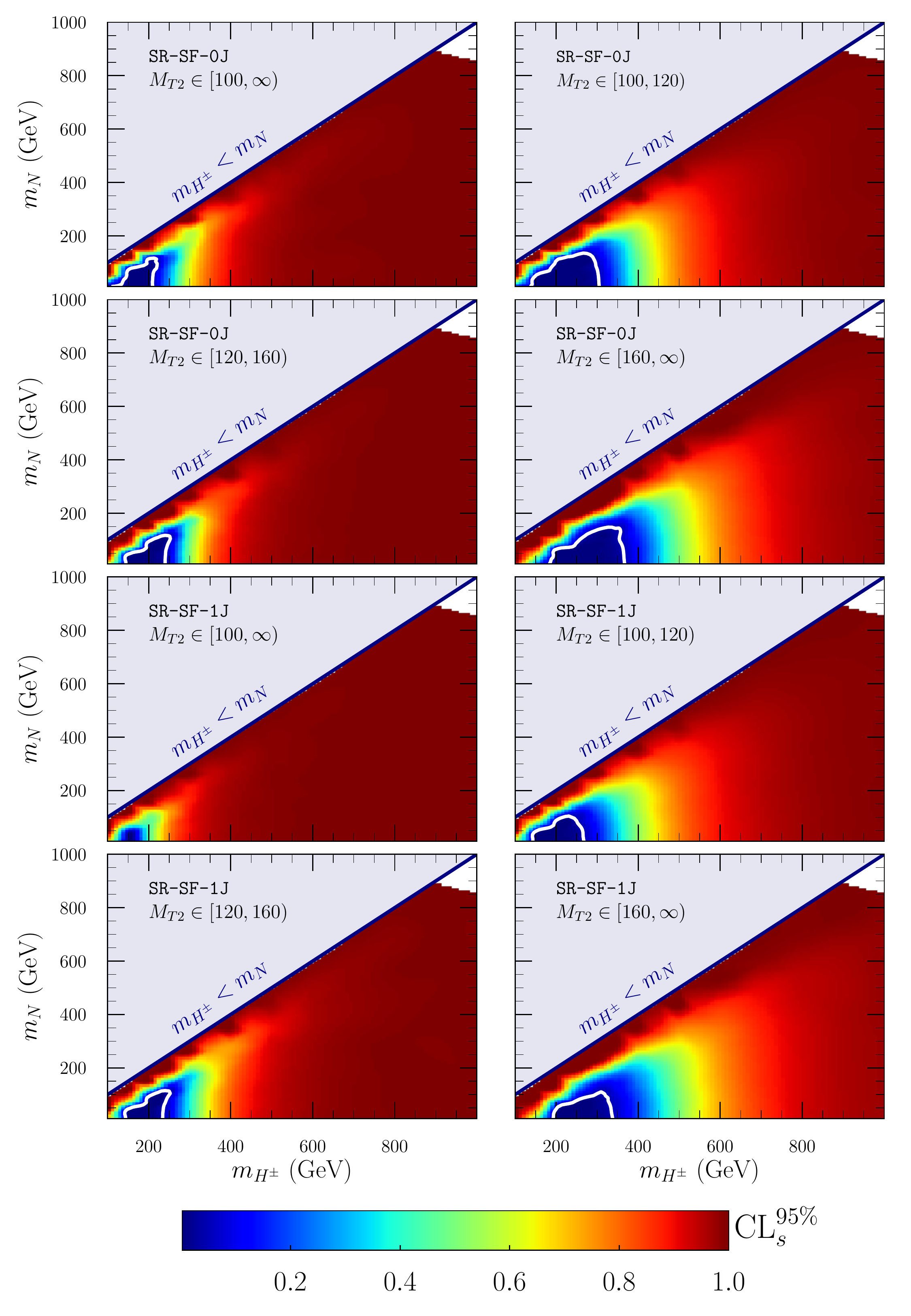} 
\caption{The CL$_s$ at $95\%$ CL projected on the mass of the charged Higgs mass ($m_{H^\pm}$) and the mass of the Majorana DM ($m_N$) for different signal regions for the SR-SF-0J and SR-SF-1J categories. The exclusions are shown for different Inclusive SRs which depend on the range of the $M_{T2}$ variable.  The shaded light navy area is forbidden by the constraint $m_{H^\pm} > m_{N}$ required by DM stability.}
\label{fig:CLs_SR-SF}
\end{figure}

\begin{table}[!t]
\setlength\tabcolsep{11pt}
{\begin{tabular}{@{}ccccc@{}} \toprule \toprule
Region         & SR-SF-0J            & SR-SF-0J            & SR-SF-0J           & SR-SF-0J \\
$M_{T2}~$(GeV) & $\in [100,\infty)$  & $\in [160,\infty)$  & $\in [100,120)$    & $\in [120,160)$ \\
\toprule
Backgrounds (Linear) & $3107\pm258$ & $805\pm64$ & $1208\pm129$ & $1100\pm107$ \\
\toprule
Backgrounds (Poisson) & $3107\pm55$ & $805\pm13$ & $1208\pm27$ & $1100\pm23$ \\ \toprule
$m_{H^\pm}, m_{N} = (500, 10)~$GeV   & $125.2$ & $105.7$ & $6.5$ & $12.9$ \\
$m_{H^\pm}, m_{N} = (500, 425)~$GeV  & $12.9$  & $0.06$ & $8.6$ & $4.3$\\
\toprule \toprule
Region         & SR-SF-1J            & SR-SF-1J            & SR-SF-1J           & SR-SF-1J \\
$M_{T2}~$(GeV) & $\in [100,\infty)$  & $\in [160,\infty)$  & $\in [100,120)$    & $\in [120,160)$ \\
\toprule
Backgrounds (Linear) & $2676\pm259$ & $777\pm108$ & $1036\pm172$ & $863\pm86$ \\
\toprule
Backgrounds (Poisson) & $2676\pm55$ & $777\pm23$ & $1036\pm37$ & $863\pm18$ \\
\toprule
$m_{H^\pm}, m_{N} = (500, 10)~$GeV & $88.5$ & $75.5$ & $4.3$ & $8.6$ \\
$m_{H^\pm}, m_{N} = (500, 425)~$GeV & $8.6$ & $0.2$ &  $4.3$ & $4.3$ \\
\toprule \toprule
\end{tabular}
\caption{Expected background event yields, at $\mathcal{L} = 3000~\mathrm{fb}^{-1}$, for the `same flavor' (SF) inclusive signal regions SRs estimated by extrapolating the results in Table \ref{tab:NeventsLHC}. The uncertainties on the background yields are computed by extrapolating the errors at the LHC$_{139 {\rm fb}^{-1}}$ linearly (assuming they are dominated by systematic uncertainties) or according to a Poisson distribution (assuming they are dominated by statistics). For reference, we show the number of events in the SRs for few benchmark points in our model corresponding to $(m_{H^\pm}, m_N) =(500, 10),~\mathrm{and}~(500, 425)~$GeV.}
 \label{tab:NeventsHL-LHC}}
\end{table}

The exclusion limits on the parameter space based on the CL$_{s}$ prescription are obtained using \texttt{MadAnalysis}. Initially, the signal event number passing the selection of a given signal region can be estimated as follows:
\begin{eqnarray}
N_{\rm s} = \epsilon_{\rm s}\ {\cal L}\ \sigma,
\end{eqnarray}
where $\cal L$ represents the recorded integrated luminosity ($139~$fb$^{-1}$), $\epsilon_{\rm s}$ is the cumulative efficiency within the SR, and $\sigma = \sigma({p p \to H^+ H^- jj}) \times {\rm BR}^2_{H^\pm \to \ell^\pm N}$  is the merged production cross section  of the Charged Higgs pairs. 
For each defined signal region, the observed event yield ($n_{\rm obs)}$, the fitted background yields ($n_{\rm b})$, and the uncertainty on the background predictions ($\Delta n_{\rm b})$  are provided (see Table \ref{tab:NeventsLHC} for details). A large number of toy experiments, denoted by $N_{\rm toys}$, is generated where in each round the $p_{\rm b}$ and $p_{{\rm b} + {\rm s}}$ probabilities are calculated following the steps:

\begin{itemize}
    \item The expected number of background events ($N_b$) is generated randomly from a Gaussian distribution where its mean is $N_{\rm b}$ and its standard deviation is $\Delta n_{\rm b}$. The actual background events number ($\tilde{N}_{\rm b}$) is obtained by using $N_b$ as parameter from Poisson distribution. Hence, the background probability $p_{\rm b}$ reads:
    
    \begin{eqnarray}
     p_{\rm b} = \frac{N_{\rm toys}(\tilde{N}_{\rm b} \leq n_{\rm obs})}{N_{\rm toys}},
     \label{eq:PB}
    \end{eqnarray}
    where $N_{\rm toys}(\tilde{N}_{\rm b} \leq n_{\rm obs})$\footnote{We note that only the number of toy experiments yielding positive $N_b$ are retained.} illustrates the toy experiments number in which the condition $\tilde{N}_{\rm b} \leq n_{\rm obs}$ is verified.
    \item The second step consists of computing the signal-plus-background probability $p_{{\rm b} + {\rm s}}$. To get the latter probability, the actual number of signal-plus-background events ($\tilde{N}_{\rm s} + \tilde{N}_{\rm b}$) is generated randomly following a Poisson distribution using the $n_s + N_b$ parameter. Therefore, in this case the probability, $p_{{\rm b} + {\rm s}}$, can be defined by
    \begin{eqnarray}
     p_{{\rm b} + {\rm s}} = \frac{N_{\rm toys}(\tilde{N}_{\rm s} + \tilde{N}_{\rm b} \leq n_{\rm obs})}{N_{\rm toys}},
     \label{eq:PBPlusS}
    \end{eqnarray}
    with  $N_{\rm toys}(\tilde{N}_{\rm s} + \tilde{N}_{\rm b} \leq n_{\rm obs})$ represents the number of toy experiments\footnote{We note that only the number of toy experiments yielding positive $N_b+n_s$ are retained.} in which the condition $\tilde{N}_{\rm s} + \tilde{N}_{\rm b} \leq n_{\rm obs}$ is satisfied.
\end{itemize}

The obtained values of the background probability ($p_{\rm b}$) and and the signal-plus-background probability ($ p_{{\rm b} + {\rm s}}$) in \ref{eq:PB} and \ref{eq:PBPlusS}, respectively, are then used to compute the CL$_s$ which reads as
\begin{eqnarray}
{\rm CL}_s = {\rm max}\bigg(0, 1-\frac{p_{{\rm b}+{\rm s}}}{p_{\rm b}}\bigg).
\end{eqnarray}
 
In Fig. \ref{fig:CLs_SR-SF}, we display the exclusion contours projected on the charged Higgs mass ($m_{H^\pm}$), and the Majorana DM mass, ($m_N$), for all inclusive signal regions.  It can clearly be seen that the SR-SF-0J implies more constraining power compared to SR-SF-1J category. Therefore, the strongest exclusion comes from the SR-SF-0J with $M_{T2} \in [160, \infty)$ which rules out  the charged Higgs masses up to $380~$GeV. As matter of fact, the search itself does not constrain small mass splittings; i.e. $m_{H^\pm} - m_{N} < 50~$GeV. \\

In our work we use \texttt{MadAnalysis} \cite{Araz:2019otb} to cope with these limits at the expected luminosity $\mathcal{L}=3000~\mathrm{fb}^{-1}$. First, the number of expected background events in this case is calculated as:
\begin{eqnarray}
 n_{\rm b}^{3000} = \frac{3000}{139} n_{\rm b} = 21.58~n_{\rm b}. 
\end{eqnarray}
Moreoever, the number of observed events ($n_{\rm obs}$) will be set to the extrapolated background event yields, i.e. $n_{\rm obs}^{3000} = n_{\rm b}^{3000}$. Consequently,  the extrapolated errors can be computed as:
\begin{eqnarray}
 \Delta n_{\rm b}^{3000} = 21.58~\Delta_{\rm b, \rm syst} \oplus 4.64~\Delta_{\rm b, \rm stat}.
\end{eqnarray}
Table \ref{tab:NeventsHL-LHC} lists the extrapolated event yields for the backgrounds and the signals. One should mention that the uncertainties on the background contributions are assumed to be dominated by just the statistical errors. 

\bibliographystyle{JHEP}
\bibliography{main.bib}


\end{document}